\documentclass[aps,pra,10pt,twocolumn,superscriptaddress]{revtex4-2}
\usepackage{amsmath}
\usepackage{amssymb}
\usepackage{bbm}
\usepackage{framed}
\usepackage{epstopdf}
\usepackage{color}
\usepackage[colorlinks = true, urlcolor = blue]{hyperref}
\usepackage{epstopdf}
\usepackage{dsfont}
\usepackage{amsthm}
\usepackage{wrapfig}
\usepackage{relsize}
\usepackage{bm}
\usepackage[latin1]{inputenc}
\usepackage{enumitem}
\usepackage{natbib}
\usepackage{graphicx}
\usepackage{physics}
\usepackage{xcolor}
\usepackage{hyperref}

\begin{document}

	\title{Quantum capacity analysis of multi-level amplitude damping channels}
	\author{Stefano Chessa}
	\email{stefano.chessa@sns.it}
	\author{Vittorio Giovannetti}
	\affiliation{NEST, Scuola Normale Superiore and Istituto Nanoscienze-CNR, I-56126 Pisa, Italy}
	\date{\today}

\begin{abstract}
The set of \textit{Multi-level Amplitude Damping} (MAD) quantum channels is introduced as a generalization of the standard qubit Amplitude Damping Channel to 
quantum systems of finite dimension $d$. In 
the special  case of 
 $d=3$, by exploiting degradability, data-processing inequalities, and channel isomorphism, we compute the associated quantum and private
 classical capacities for a rather wide class of maps, extending the set of solvable models known so far.  We proceed then to the evaluation of the entanglement assisted, quantum and classical, capacities.
\end{abstract}

\maketitle

\section{Introduction} \label{sec.Intro} 
The main goal of quantum information and communication theory is to understand how can we store, process and transfer information in a reliable way and, from the physical point of view, to individuate realistic platforms by means of which performing these tasks. All by exploiting the characteristic features of quantum mechanics. Focusing on quantum communication, every communication protocol can be seen as a physical system (the encoded message) undergoing some physical transformation that translates it in space or time. Any real-world application though suffers from some kind of noise, each of which can be in turn described as a quantum process or equivalently as a quantum channel. Following the work of Shannon \cite{SHANNON} and the later quantum generalizations, the ability of a quantum channel to preserve the encoded classical or quantum information is described by its capacities~\cite{HOLEGIOV,BENNETTSHOR}. In the classical case we can only transfer classical information, hence we only need to deal with the classical capacity. In the quantum framework we can also transfer quantum states and consequently, in addition to the classical capacity, we count also the quantum capacity. Moreover, the family of capacities associated with a quantum channel can be enlarged assuming the communicating parties to be able to perform specific tasks or to share further resources such as, for instance, entanglement \cite{HOLEVO BOOK,WILDE,WATROUSBOOK,HOLEGIOV,NC,SURVEY}. 

In this paper we will focus on the specific and well known model for quantum noise given by the amplitude damping channel (ADC). While the ADC has been thoroughly studied and characterized, in terms of capacities in various settings, for the qubit framework \cite{QUBIT ADC, DARRIGO, QUBIT ADC 1, QUBIT ADC 2}, a general treatise for qudit ($d$-dimensional) systems is still missing and likely not possible to attain. Because of these reasons ADC for $d>2$ has to be approached case by case, and the literature regarding capacities of fixed finite dimensions ADC is still remarkably short \cite{QUBIT ADC 2,QUDIT,QUDIT1}. Our interest in the topic is due to the fact that higher dimensional systems have attracted the attention of a growing number of researchers in recent years, since they have been shown to provide potential advantages both in terms of computation (see e.g. \cite{COMP,COMP2,COMP1,COMP3,COMP4,COMP5}) and communication or error correction (see e.g. \cite{COMM,COMM1,COMM2,COMM3}) together with the fact that more experimental implementations have been progressively made available (see e.g. \cite{EXP,EXP1,EXP2,EXP3,EXP4,EXP5,EXP6,EXP7}). Among non-qubit systems, three-dimensional systems (qutrit) have received particular consideration because of their relative accessibility both theoretically and experimentally (see e.g. \cite{QUTRIT,QUTRIT1,QUTRIT2,QUTRIT3,QUTRIT4,QUTRIT5,QUTRIT6,QUTRIT7,QUTRIT8,QUTRIT9}).
In addition to that, new results on the quantum capacity of finite dimensional channels can also be applied to higher dimensional maps via the \textit{Partially Coherent Direct Sum} (PCDS) channels approach \cite{ARTICOLO1}, placing in a wider context the efforts dedicated to the analysis of non-qubit channels. Considering this, we will start a first systematic analysis of the ADC on the qutrit space: while we will not approach the issue of the classical capacity of the channel, we will focus on the quantum capacity, private classical capacity and entanglement assisted capacities, trying to understand in which conditions these quantities can be known. \\

The paper is structured as follows. In Sec.~\ref{sec.Model} we introduce the model and notations we used for the qutrit MAD. In Sec.~\ref{sec:quantum capacity} we proceed to the study of the quantum capacity and private classical capacity of the qutrit MAD in various configurations. In Sec.~\ref{sec:Ent ass} we repeat the same analysis for the entanglement assisted quantum and classical capacities. 

\section{Settings} \label{sec.Model}

The transformations  we focus on in the present work are special instances of the 
 multi-level versions of the qubit ADC~\cite{QUBIT ADC}, hereafter indicated as MAD channels in brief, which effectively 
describe the decaying of energy levels of  a $d$-dimensional quantum system A. In its most general  form, given $\{|i\rangle\}_{i=0, \cdots, d-1}$ an orthonormal basis of the Hilbert space 
${\cal H}_{\text{A}}$ associated with A (hereafter dubbed the computational basis of the problem),  a MAD channel ${\cal D}$ is a Completely Positive Trace Preserving (CPTP) mapping~\cite{HOLEVO BOOK,WILDE,WATROUSBOOK,HOLEGIOV,NC,SURVEY}    acting on the set $\mathcal{L}(\mathcal{H}_{\text{A}})$ of linear operators of the system,
 defined by the following set of $d(d-1)/2+1$
Kraus operators
 \begin{eqnarray}\nonumber 
\hat{K}_{ij}&\equiv&\sqrt{\gamma_{ji}}\ket{i}\!\!\bra{j}, \qquad  \mbox{$\forall~i,j$ {s.t.}  $0\leq i<j\leq d-1$,}  \\  
 \label{K0} 
\hat{K}_0 &\equiv&  \ket{0}\!\!\bra{0} + 
\sum\limits_{1\leq j\leq d-1} \sqrt{1 -\xi_j} \ket{j}\!\!\bra{j}\;, \label{eq.Kraus}
 \end{eqnarray} 
with  $\gamma_{ji}$ real quantities describing the decay rate from the $j$-th to the $i$-th level that fulfill the conditions
\begin{eqnarray} \label{BIGMA} 
\left\{ \begin{array}{l} 
0\leq \gamma_{ji} \leq 1 \;, \qquad \mbox{$\forall~i,j$ {s.t.}  $0\leq i<j\leq d-1$,} \\\\
\xi_j\equiv \sum\limits_{0\leq i <j} \gamma_{ji} \leq 1\;,  \qquad \forall j= 1,\cdots, d-1\;.
\end{array} \right.
\end{eqnarray} 
\begin{figure}[t!]
  \includegraphics[width=\linewidth]{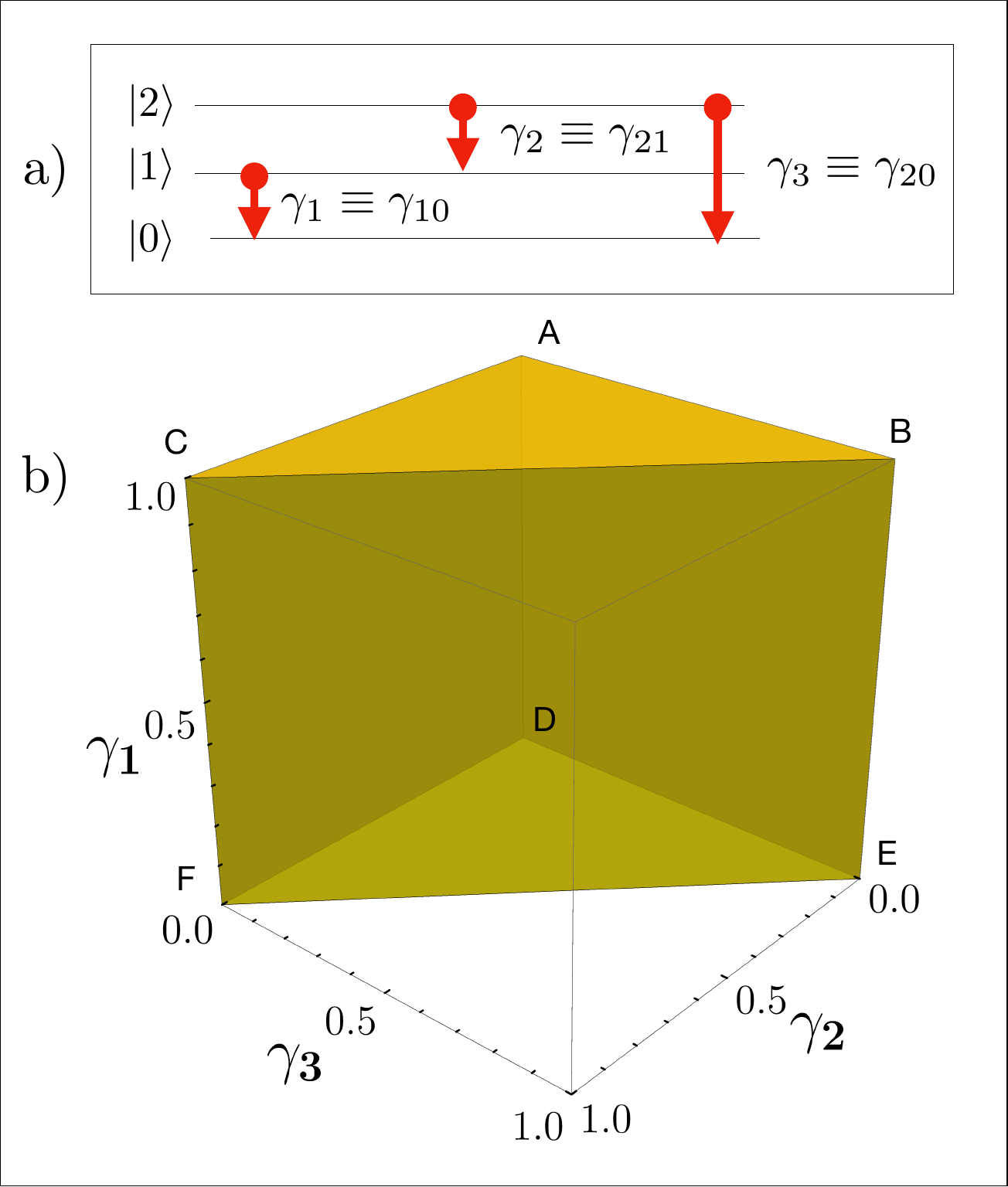}
  \caption{Top panel: schematic representation of the action of the MAD channel    $\mathcal{D}_{\vec{\gamma}}$ on a 3-level system.
Bottom panel: the admitted region of the damping parameters space: the transformation is CPTP if and only if the rate vector $\vec{\gamma}$ 
  belongs to the yellow region defined in  Eq.~(\ref{REGION}).}
  \label{fig:decay}
\end{figure}
Accordingly, given $\hat{\rho}\in \mathfrak{S}({\cal H}_A)$ a generic density matrix of the system A, the MAD channel ${\cal D}$ will transform it into the output
state defined as 
\begin{eqnarray}
\mathcal{D}(\hat{\rho})&=&\hat{K}_0\hat{\rho} \hat{K}_0^{\dagger}+\sum\limits_{0\leq i<j\leq d-1}\hat{K}_{ij}\hat{\rho} \hat{K}_{ij}^{\dagger}\;, \nonumber \\ \label{MAD} 
&=&\hat{K}_0\hat{\rho} \hat{K}_0^{\dagger}+\sum\limits_{0\leq i<j\leq d-1}\gamma_{ji}
|i\rangle\langle i|\;  \langle j|\hat{\rho} |j\rangle\;.
\end{eqnarray}
By construction ${\cal D}$ always admits the ground state $|0\rangle$ as a fixed point, i.e. $
\mathcal{D}(|0\rangle\langle 0|)=|0\rangle\langle 0|$, 
even though, depending on the specific values of the coefficients $\gamma_{ji}$, other input states may fulfill the same property as well. 
Limit cases are $\gamma_{ji}=0\; \forall\, i, \, j$, where all levels are untouched and ${\cal D}$  reduces to the noiseless identity channel  $\text{Id}$ which preserves all the input states of A. On the opposite extreme are those examples in which for some $j$ we have  $\xi_j=1$, corresponding to the scenario where the $j$-th level becomes totally depopulated at the end of the transformation. 
The maps~(\ref{MAD}) 
 provide also a natural playground to describe Partially Coherent
Direct Sum (PCDS) channels~\cite{ARTICOLO1}. 
Last but not the least, an important and easy to verify  property of the maps~(\ref{MAD})
 is that they are covariant under
the group formed by the unitary transformations $\hat{U}$ which are diagonal in the 
computational basis $\{|i\rangle\}_{i=0, \cdots, d-1}$, i.e. 
\begin{eqnarray}\label{COV} 
{\cal D}(\hat{U} \hat{\rho} \hat{U}^\dag) = \hat{U} {\cal D}( \hat{\rho} )\hat{U}^\dag\;,
\end{eqnarray} 
for all inputs $\hat{\rho}$.

For what concerns the present work, we shall restrict our analysis to the special 
set of MAD channels~(\ref{MAD}) associated with  a qutrit system ($d=3$) whose
decay processes, pictured in  the top panel of Fig.~\ref{fig:decay}, are fully characterized by only 
three rate parameters $\gamma_{ji}$  that for the ease of notation we rename with the cartesian components of a  3D vector $\vec{\gamma}\equiv( \gamma_{1}, \gamma_{2}, \gamma_{3})$. Accordingly, expressed in terms of the 
matrix representation induced by the computational basis $\{ |0\rangle, |1\rangle, |2\rangle\}$, the Kraus operators~(\ref{eq.Kraus}) write explicitly as 
\begin{equation}\label{eq:Krausnewee}
\begin{split}
&\small{\hat{K}_0=\begin{pmatrix}
1 & 0 & 0\\
0 & \sqrt{1-\gamma_1} & 0\\
0 & 0 & \sqrt{1-\gamma_2-\gamma_3}
\end{pmatrix},\quad
\hat{K}_{01}=\begin{pmatrix}
0 & \sqrt{\gamma_1} & 0\\
0 & 0 & 0\\
0 & 0 & 0
\end{pmatrix}},\\
& \small{ \hat{K}_{12}=\begin{pmatrix}
0 & 0 & 0\\
0 & 0 & \sqrt{\gamma_2}\\
0 & 0 & 0
\end{pmatrix},\quad
\hat{K}_{03}=\begin{pmatrix}
0 & 0 & \sqrt{\gamma_3}\\
0 & 0 & 0\\
0 & 0 & 0
\end{pmatrix}},
\end{split}
\end{equation}
with CPTP conditions~(\ref{BIGMA}) given by
 \begin{eqnarray}  \label{REGION} 
\left\{ \begin{array}{l}
0\leq \gamma_j \leq 1 \;, \qquad \qquad \forall j=1,2,3\;, \\ \\
 \gamma_2 + \gamma_3 \leq 1\;,
\end{array} \right.
\end{eqnarray} 
which produce the volume visualized in the bottom panel of Fig.~\ref{fig:decay}.

The resulting mapping~(\ref{MAD}) for the channel $\mathcal{D}_{( \gamma_{1}, \gamma_{2}, \gamma_{3}) }$ 
reduces hence to the following expression 
\begin{widetext}
\begin{equation}\label{eq:matrix expression}
\mathcal{D}_{\vec{\gamma}}(\hat{\rho})=\begin{pmatrix}
\rho_{00}+\gamma_1 \rho_{11}+\gamma_3 \rho_{22}& \sqrt{1-\gamma_1}\rho_{01} & \sqrt{1-\gamma_2-\gamma_3}\rho_{02}\\
\sqrt{1-\gamma_1}\rho_{01}^* & (1-\gamma_1)\rho_{11}+\gamma_2 \rho_{22} & \sqrt{1-\gamma_1}\sqrt{1-\gamma_2-\gamma_3}\rho_{12}\\
\sqrt{1-\gamma_2-\gamma_3}\rho_{02}^* & \sqrt{1-\gamma_1}\sqrt{1-\gamma_2-\gamma_3}\rho_{12}^* & (1-\gamma_2-\gamma_3)\rho_{22}
\end{pmatrix},
\end{equation}
\end{widetext}
while the associated complementary 
CPTP transformation~\cite{HOLEVO BOOK,WILDE,WATROUSBOOK,HOLEGIOV} computed as in Eq.~(\ref{eq:app compl chan}) of   Appendix~\ref{MATHAPP},
for generic choices of the system parameters, transforms A into a 4-dimensional state via the mapping  
\begin{widetext}
\begin{equation}\label{eq:matr expr compl}
\tilde{\mathcal{D}}_{\vec{\gamma}}(\hat{\rho})=\left(
\begin{array}{cccc}
\rho_{00}+(1 -\gamma_1)\rho_{11}+ (1-\gamma_2-\gamma_3 ) \rho_{22} & \sqrt{\gamma _1}  \rho_{01} & \sqrt{1-\gamma_1} \sqrt{\gamma_2} \rho_{12} & \sqrt{\gamma_3} \rho_{02} \\
 \sqrt{\gamma_1} \rho_{01}^* & \gamma_1 \rho_{11} & 0 & \sqrt{\gamma_1} \sqrt{\gamma_3} \rho_{12} \\
 \sqrt{1-\gamma_1} \sqrt{\gamma_2} \rho_{12}^* & 0 & \gamma_2 \rho_{22} & 0 \\
 \sqrt{\gamma_3} \rho_{02}^* & \sqrt{\gamma_1} \sqrt{\gamma_3} \rho_{12}^* & 0 & \gamma_3 \rho_{22} \\
\end{array}
\right)\;, 
\end{equation}
\end{widetext}
where for $i,j \in 0,1,2$, $\rho_{ij} \equiv \langle i|\hat{\rho} |j\rangle$ are the  matrix entries of the input density operator $\hat{\rho} \in \mathfrak{S}({\cal H}_{\text{A}})$.

\subsection{Composition rules}\label{sec:comp} 

It is relatively easy to verify that the set of qutrit MAD channels~(\ref{eq:matrix expression}) is close under concatenation.
Specifically we notice that given  $\mathcal{D}_{\vec{\gamma}'}$ and $\mathcal{D}_{\vec{\gamma}''}$ 
with $\vec{\gamma}''=(\gamma''_1,\gamma''_2,\gamma''_3)$ and $\vec{\gamma}'=(\gamma'_1,\gamma'_2,\gamma'_3)$
two rate vectors fulfilling the conditions~(\ref{REGION}), we have 
\begin{equation}\label{eq:composition}
\mathcal{D}_{\vec{\gamma}'}\circ\mathcal{D}_{\vec{\gamma}''}=\mathcal{D}_{\vec{\gamma}}\;, 
\end{equation}
with $\vec{\gamma}=(\gamma_1,\gamma_2,\gamma_3)$  a new rate vector of components 
\begin{equation} \label{GAMMACOM} 
\begin{cases}
\gamma_1=\gamma''_1+\gamma'_1-\gamma'_1 \gamma''_1\;,\\
\gamma_2=\gamma''_2(1-\gamma'_1-\gamma'_2)+\gamma'_2(1-\gamma''_3)\;, \\
\gamma_3=\gamma''_3+\gamma''_2(\gamma'_1-\gamma'_3)+\gamma'_3(1-\gamma''_3)\;.
\end{cases}
\end{equation}
which also satisfies~(\ref{REGION})
(hereafter we shall use the symbol ``$\circ$"  to represent super-operator composition). 
The importance of Eq.~(\ref{eq:composition}) for the problem we are facing stems from 
channel data-processing inequalities  (or bottleneck) inequalities~\cite{KEYL,WILDE1,NC}, according to which,
any information capacity functional $\Gamma$~\cite{HOLEGIOV} such as  the quantum capacity $Q$, the classical capacity $C$, the private classical capacity $C_p$, the entanglement 
assisted classical capacity $C_E$ etc., computed for a CPTP map $\Phi=\Phi'\circ \Phi''$ obtained by concatenating channel $\Phi'$ with channel $\Phi''$,  must fulfill the following relation
\begin{equation}\label{eq:data processing}
\Gamma(\Phi)\leq \min\{ \Gamma(\Phi'),  \Gamma(\Phi'')\} \;.
\end{equation}
 Applied to Eq.~(\ref{eq:composition}), the above inequality can be used to predict monotonic behaviors for the capacity  $\Gamma(\mathcal{D}_{\vec{\gamma}})$
 as a function of the rate vector $\vec{\gamma}$, that allows us to provide useful lower and upper bounds which in some case permit to extend the capacity formula to domain where other techniques (e.g. degradability analysis)
 fail. 
In particular we notice that for single-decay MAD channels where only one component of the rate vector is different from zero (say $\gamma_1$) 
we get 
\begin{equation} \label{COMBO} 
\mathcal{D}_{(\gamma^{'}_1,0,0)} \circ \mathcal{D}_{(\gamma^{''}_1,0,0)}= \mathcal{D}_{(\gamma^{''}_1,0,0)} \circ \mathcal{D}_{(\gamma^{'}_1,0,0)}  = \mathcal{D}_{(\gamma_1,0,0)}\;, 
\end{equation} 
with  $\gamma_1$ as in the first identity of Eq.~(\ref{GAMMACOM}). Accordingly we can conclude that all the capacities $\Gamma(\mathcal{D}_{(\gamma_1,0,0)})$ 
should be non increasing functionals of the parameter $\gamma_1$, i.e. 
\begin{eqnarray} \label{DECGM} 
\Gamma(\mathcal{D}_{(\gamma_1,0,0)})\geq \Gamma(\mathcal{D}_{(\gamma_1',0,0)}), \qquad \forall \gamma_1\leq \gamma_1'\;,
\end{eqnarray} 
(the same expressions and conclusions apply also for $\mathcal{D}_{(0,\gamma_2,0)}$ and $\mathcal{D}_{(0,0,\gamma_3)}$). 
Composing single-decay MAD channels characterized by rate vectors pointing along different cartesian axis, in general creates  
maps    with higher rank 
of the resulting vector rate. Specifically from Eq.~(\ref{eq:composition}) it follows 
 that, for an arbitrary choice of the rate vector $\vec{\gamma}={(\gamma_1,\gamma_2,\gamma_3)}$ in the allowed CPTP domain the MAD channel $\mathcal{D}_{(\gamma_1,\gamma_2,\gamma_3)}$  can be expressed as 
\begin{eqnarray}\label{eq:compos 3dec 2}
\mathcal{D}_{(\gamma_1,\gamma_2,\gamma_3)} &=& \mathcal{D}_{(0,0,\bar{\gamma}_3)}\circ\mathcal{D}_{(0,\gamma_2,0)}\circ\mathcal{D}_{(\gamma_1,0,0)}  \\
&=&\mathcal{D}_{(0,\bar{\gamma}_2,0)}\circ \mathcal{D}_{(0,0,\gamma_3)} \circ\mathcal{D}_{(\gamma_1,0,0)} \;, \label{eq:compos 3dec 2new}
\end{eqnarray}
with 
\begin{eqnarray} 
\bar{\gamma}_3\equiv \frac{\gamma_3}{1-\gamma_2}\;,  \qquad \quad 
\bar{\gamma}_2\equiv \frac{\gamma_2}{1-\gamma_3}\;,  
\end{eqnarray} 
which because of the constraint~(\ref{REGION}) are  properly defined rates.
As a direct consequence  of Eqs.~(\ref{eq:data processing}) and ~(\ref{COMBO}) it then follows that the capacities 
$\Gamma(\mathcal{D}_{(\gamma_1,\gamma_2,\gamma_3)})$
must be non-increasing functionals of all the cartesian components of rate vector $\vec{\gamma}$, i.e. 
\begin{equation} \label{DECGMstrong} 
\Gamma(\mathcal{D}_{(\gamma_1,\gamma_2,\gamma_3)})\geq \Gamma(\mathcal{D}_{(\gamma'_1,\gamma'_2,\gamma'_3)} ), \quad \forall \gamma'_i\geq \gamma_i\;.
\end{equation} 
and must be restricted by the upper bound 
\begin{equation}\label{eq:boundCOMBO}
\Gamma(\mathcal{D}_{(\gamma_1,\gamma_2,\gamma_3)}) \leq \min\{ \Gamma( \mathcal{D}_{(\gamma_1,0,0)}),\mathcal{D}_{(0,\bar{\gamma}_2,0)}),\mathcal{D}_{(0,0,\bar{\gamma}_3)})\}\;.
\end{equation}
As a further refinement notice that, setting $\gamma_2=0$ in Eqs.~(\ref{eq:compos 3dec 2}) and (\ref{eq:compos 3dec 2new})
we get
\begin{equation} 
\mathcal{D}_{(\gamma_1,0,\gamma_3)} =\mathcal{D}_{(\gamma_1,0,0)}\circ\mathcal{D}_{(0,0,\gamma_3)}=\mathcal{D}_{(0,0,\gamma_3)}\circ \mathcal{D}_{(\gamma_1,0,0)}\;,
\end{equation} 
which replaced back into  Eq.~(\ref{eq:compos 3dec 2new}) gives us 
\begin{equation}\label{eq:compos 3dec}
\mathcal{D}_{(\gamma_1,\gamma_2,\gamma_3)} = \mathcal{D}_{(0,\bar{\gamma}_2,0)}\circ\mathcal{D}_{(\gamma_1,0,\gamma_3)}\;,
\end{equation}
which allows us to replace~(\ref{eq:boundCOMBO}) with the stronger requirement 
\begin{equation}\label{eq:boundCOMBO1}
\Gamma(\mathcal{D}_{(\gamma_1,\gamma_2,\gamma_3)}) \leq \min\{ \Gamma( \mathcal{D}_{(0,\bar{\gamma}_2,0)}), \Gamma(\mathcal{D}_{(\gamma_1,0,\gamma_3)})\}\;.
\end{equation}
Similarly by setting $\gamma_1=0$ we get 
\begin{equation} 
\mathcal{D}_{(0,\gamma_2,\gamma_3)} =\mathcal{D}_{(0,0,\bar{\gamma}_3)}\circ\mathcal{D}_{(0,\gamma_2,0)}=\mathcal{D}_{(0,\bar{\gamma}_2,0)}\circ \mathcal{D}_{(0,0,\gamma_3)}\;,
\end{equation} 
that yields
\begin{eqnarray}\label{eq:compos 3decnew1}
\mathcal{D}_{(\gamma_1,\gamma_2,\gamma_3)} &=& \mathcal{D}_{(0,\gamma_2,\gamma_3)}\circ\mathcal{D}_{(\gamma_1,0,0)}\;, \end{eqnarray}
and 
\begin{equation}  \label{eq:boundCOMBO2} 
\Gamma(\mathcal{D}_{(\gamma_1,\gamma_2,\gamma_3)}) \leq \min\{ \Gamma(\mathcal{D}_{(0,\gamma_2,\gamma_3)}), \Gamma(\mathcal{D}_{(\gamma_1,0,0)})\}\;.
\end{equation}
Finally setting $\gamma_3=0$ in Eqs.~(\ref{eq:compos 3dec 2})
we get 
\begin{equation} 
\mathcal{D}_{(\gamma_1,\gamma_2,0)} =\mathcal{D}_{(0,\gamma_2,0)} \circ \mathcal{D}_{(\gamma_1,0,0)}\;, \end{equation} 
that leads to 
\begin{eqnarray}\label{eq:compos 3decnew13}
\mathcal{D}_{(\gamma_1,\gamma_2,\gamma_3)} &=& \mathcal{D}_{(0,0,\bar{\gamma}_3)}\circ\mathcal{D}_{(\gamma_1,\gamma_2,0)} \;, \end{eqnarray}
and
\begin{equation}  \label{eq:boundCOMBO3} 
\Gamma(\mathcal{D}_{(\gamma_1,\gamma_2,\gamma_3)}) \leq \min\{ \Gamma(\mathcal{D}_{(\gamma_1,\gamma_2,0)}), \mathcal{D}_{(0,0,\bar{\gamma}_3)}\}\;.
\end{equation}

\section{Quantum and private classical capacities for qutrit MAD}\label{sec:quantum capacity}

The quantum capacity $Q$ of a quantum channel  is a measure of how faithfully quantum states
can be transmitted  from the input to the output of the associated CPTP map by exploiting proper  encoding and decoding
procedures that act on multiple transmission stages~\cite{HOLEVO BOOK,WILDE,WATROUSBOOK,HOLEGIOV,NC,SURVEY}. 
The private classical capacity $C_p$ instead quantifies 
 the amount of classical information transmittable per channel use under the extra requirement that 
 the entire signaling process allows the communicating parties to be protected by eavesdropping by an adversary agent that is controlling the communication line.
The explicit evaluation of these important functionals is one of the most elusive task of quantum information theory, as testified
by the limited number of examples which allow for an explicit solution. 
 For a comprehensive, self-consistent  introduction to the technical problems involved in this calculation we refer the reader 
to the Appendix~\ref{MATHAPP}, where we present   the notions of complementary channel, coherent information,
and degradability and where we introduce
the explicit functionals~\cite{QCAP1,QCAP2,QCAP3} we need to optimize. 
Building up from these premises here  we present a thoughtful 
characterization of the quantum capacity $Q({\cal D}_{\vec{\gamma}})$ and the private classical 
capacity $C_p({\cal D}_{\vec{\gamma}})$ of the qutrit MAD channel~${\cal D}_{\vec{\gamma}}$ defined in Eq.~(\ref{eq:matrix expression}).
We stress that while failing to provide the explicit solution  for all rate vectors $\vec{\gamma}$ in the allowed domain defined by Eq.~(\ref{REGION}), in what follows
we  manage to deliver the exact values of $Q({\cal D}_{\vec{\gamma}})$  and $C_p({\cal D}_{\vec{\gamma}})$ for a quite a large class of qutrit MAD channels by  making use of degradability 
properties~\cite{DEGRADABLE}, data-processing (or bottleneck) inequalities~\cite{KEYL,WILDE1}, and channel
isomorphism.
In particular we anticipate here that, for those  ${\cal D}_{\vec{\gamma}}$ which are provably degradable~\cite{DEGRADABLE}, 
 we shall exploit the covariance property~(\ref{COV}) 
to further simplify the single-letter formula~(\ref{eq:QCapacity1}) as 
\begin{equation}
\label{eq:QCapacity1GasDIAG3}
Q({\cal D}_{\vec{\gamma}})=C_p({\cal D}_{\vec{\gamma}})=
\max_{\hat{\rho}_{\text{diag}}} \left\{ S({\cal D}_{\vec{\gamma}}(\hat{\rho}_{\text{diag}})) -
 S(\tilde{\cal D}_{\vec{\gamma}}(\hat{\rho}_{\text{diag}}))\right\} \;, 
\end{equation}
where $S(\cdots)$ is the von Neumann entropy, and 
where the maximization is performed on input states of A which are diagonal in the computational basis of the problem,
i.e. the density matrices of the form $\hat{\rho}_{\text{diag}} =  \sum_{i=0}^2 p_i |i\rangle\langle i|$ 
with $p_0,p_1,p_2\in[0,1]$ fulfilling the normalization constraint  $p_0+p_1+p_2=1$ -- 
see discussion at the end of Appendix~\ref{sec:Quant and Priv} for details. 
Notably, when applicable, Eq.~(\ref{eq:QCapacity1GasDIAG3}) relies on an optimization of a functional of only two
real variables (namely the populations $p_0$ and $p_1$) which can be easily carried on (at least numerically).

To begin with, observe that,
as anticipated in Eq.~(\ref{eq:matr expr compl}), the complementary map $\tilde{\mathcal{D}}_{\vec{\gamma}}$ of a generic qutrit MAD channel ${\cal D}_{\vec{\gamma}}$ sends the input states of A into a 4-dimensional ``environment state''. In the end this is a consequence of the fact
that the (minimal) number of Kraus operators we need to express~(\ref{eq:matrix expression})  is 4. Unfortunately this number also ensures us that the channel is not degradable: it has been indeed shown~\cite{PROP DEGR} that a necessary condition
for  any CPTP map with output dimension $3$ to be degradable is that its associated Choi rank, and consequently the minimal number of Kraus operators we need to express such transformation, is at most 3. 
This brings us to consider some simplification in the problem, e.g. by fixing some of the values of the damping parameters. One approach is represented by the selective suppression of one (or two) of the decaying channels, i.e. imposing one (or two) of the parameters $\gamma_i$ equal to 0, which we will do in Secs.~\ref{sec:single decay},~\ref{sec:double decay}, and ~\ref{sec:double decay1}. For each of these subclasses of channels we'll give a characterization, when possible, in terms of degradability, antidegradability and quantum capacity. 
A second approach that we adopt in Secs.~\ref{sec:first} and~\ref{sec:double decayplane}, consists instead to fix one of the damping parameters to its maximum allowed value, a choice that as we shall see,
will effectively allow us to reduce  the number of  degrees of freedom of the problem.

\subsection{Single-decay qutrit MAD channels}\label{sec:single decay}

We consider here instances of the qutrit MAD channel in which only one of the  three damping parameters 
$\gamma_i$ is explicitly different from zero, i.e. the maps  $\mathcal{D}_{(\gamma_1,0,0)}$,  $\mathcal{D}_{(0,\gamma_2,0)}$, and  $\mathcal{D}_{(0,0,\gamma_3)}$ associated respectively with the edges $DA$, $DF$ and $DE$ of Fig.~\ref{fig:decay}.
 It is easy to verify that these three sets of transformations can be mapped into each other via unitary conjugations that simply permute the energy levels of the system: for instance 
 $\mathcal{D}_{(0,0,\gamma_3=\gamma)}$ can be transformed into $\mathcal{D}_{(\gamma_1=\gamma,0,0)}$ by simply swapping levels $|1\rangle$ and $|2\rangle$. 
 Accordingly, as a consequence of~(\ref{eq:data processing}), 
 the capacities of these three sets must coincide, i.e. 
 \begin{equation}
 Q(\mathcal{D}_{(\gamma,0,0)}) =  Q(\mathcal{D}_{(0,\gamma,0)}) = Q(\mathcal{D}_{(0,0\gamma)}) \;, \quad \forall \gamma\in [0,1]\;, 
 \end{equation} 
 (similarly for $C_p$).
  By virtue of this fact, without loss of generality, in the following we report the analysis only for $\mathcal{D}_{(\gamma_1,0,0)}$, being the results trivially extendable to the remaining two. 
 
  It turns out that the channel $\mathcal{D}_{(\gamma_1,0,0)}$ is a special instance of the PCDS maps analyzed in Ref.~\cite{ARTICOLO1} where an explicit
  formula for $Q$ has been already derived. Still, for the sake of completeness, we find it useful to present here an alternative derivation 
  of those results which does not make explicit reference  to the PCDS structure.  
   For this purpose we observe that from Eq.~(\ref{eq:Krausnewee}) it follows that  $\mathcal{D}_{(\gamma_1,0,0)}$ possesses only two non zero Kraus operators, i.e. 
\begin{equation}\label{KRAUS121} 
\hat{K}_0=\begin{pmatrix}
1 & 0 & 0\\
0 & \sqrt{1-\gamma_1} & 0\\
0 & 0 & 1
\end{pmatrix}\quad
\hat{K}_{01}=\begin{pmatrix}
0 & \sqrt{\gamma_1} & 0\\
0 & 0 & 0\\
0 & 0 & 0
\end{pmatrix}.
\end{equation}
Transformation~(\ref{eq:matrix expression}) is then given by 
\begin{equation} \label{eq:matrixexpression1} 
\mathcal{D}_{(\gamma_1,0,0)}(\hat{\rho})=\begin{pmatrix}
\rho_{00}+\gamma_1 \rho_{11}& \sqrt{1-\gamma_1}\rho_{01} & \rho_{02}\\
\sqrt{1-\gamma_1}\rho_{01}^* & (1-\gamma_1)\rho_{11} & \sqrt{1-\gamma_1}\rho_{12}\\
\rho_{02}^* & \sqrt{1-\gamma_1}\rho_{12}^* & \rho_{22}
\end{pmatrix},
\end{equation}
and the complementary channel $\tilde{\mathcal{D}}_{(\gamma_1,0,0)}$ that can be 
expressed as a mapping that connects the system A to a 2-dimensional environmental system E, i.e. 
\begin{equation}\label{eq:compl chan D1}
\tilde{\mathcal{D}}_{(\gamma_1,0,0)}(\hat{\rho})=\begin{pmatrix}
1-\gamma_1 \rho_{11}& \sqrt{\gamma_1}\rho_{01} \\
\sqrt{\gamma_1}\rho_{01}^* & \gamma_1\rho_{11}
\end{pmatrix}.
\end{equation}
From Eq.~(\ref{eq:matrixexpression1})  it follows that, irrespectively of the value of $\gamma$, the model always 
owns a 2-dim noiseless subspace spanned by the vectors $|0\rangle$ and $|2\rangle$ ensuring a non zero lower bound for both the quantum and the private classical capacity 
 \begin{eqnarray}  \label{LOW1} 
 Q(\mathcal{D}_{(\gamma_1,0,0)}), C_p(\mathcal{D}_{(\gamma_1,0,0)}) \geq \log_2(2)=1,
 \end{eqnarray} 
which incidentally implies that  the channel $\mathcal{D}_{(\gamma_1,0,0)}$ is never anti-degradable. 
By methods discussed in Appendix~\ref{sec:Appendix compl chan}  we can also 
show that  $\mathcal{D}_{(\gamma_1,0,0)}$ is always mathematically invertible for all $\gamma_1<1$, with 
$\tilde{\mathcal{D}}_{(\gamma_1,0,0)}\circ \mathcal{D}_{(\gamma_1,0,0)}^{-1}$ CPTP for all $\gamma_1\leq \frac{1}{2}$. Accordingly, invoking~(\ref{NECANDSUFF})  we can  ensure the channel to be degradable if and only if $\gamma_1\leq \frac{1}{2}$ and use Eq.~(\ref{eq:QCapacity1GasDIAG3}) to compute its capacity value (notice that in principle the above argument leaves open 
 the possibility that the channel would be degradable also for $\gamma_1=1$, this however can be excluded by direct calculation or invoking
 the analysis of~\cite{ARTICOLO1}). 
Consequently for $\gamma_1\leq \frac{1}{2}$ we can write 
\begin{equation}
\begin{split}
&Q(\mathcal{D}_{(\gamma_1,0,0)}) =C_p(\mathcal{D}_{(\gamma_1,0,0)}) \nonumber \\
&
=\max_{p_0,p_1} \Big\{ -(1-p_0-p_1)\log_2(1-p_0-p_1)
\\
&-(p_0+\gamma_1 p_1)\log_2(p_0+\gamma_1 p_1) -(1-\gamma_1)p_1\log_2((1-\gamma_1)p_1) \\
&   +(1-\gamma_1p_1)\log_2(1-\gamma_1p_1)+\gamma_1p_1\log_2(\gamma_1 p_1)\Big\}, 
\end{split}\label{solveTHIS} 
\end{equation}
which can be solved numerically (the maximization being performed over all possible values $p_0,p_1\in [0,1]$ under the constraint that $p_0+p_1\leq 1$). 

\begin{figure}[t]
  \includegraphics[width=\linewidth]{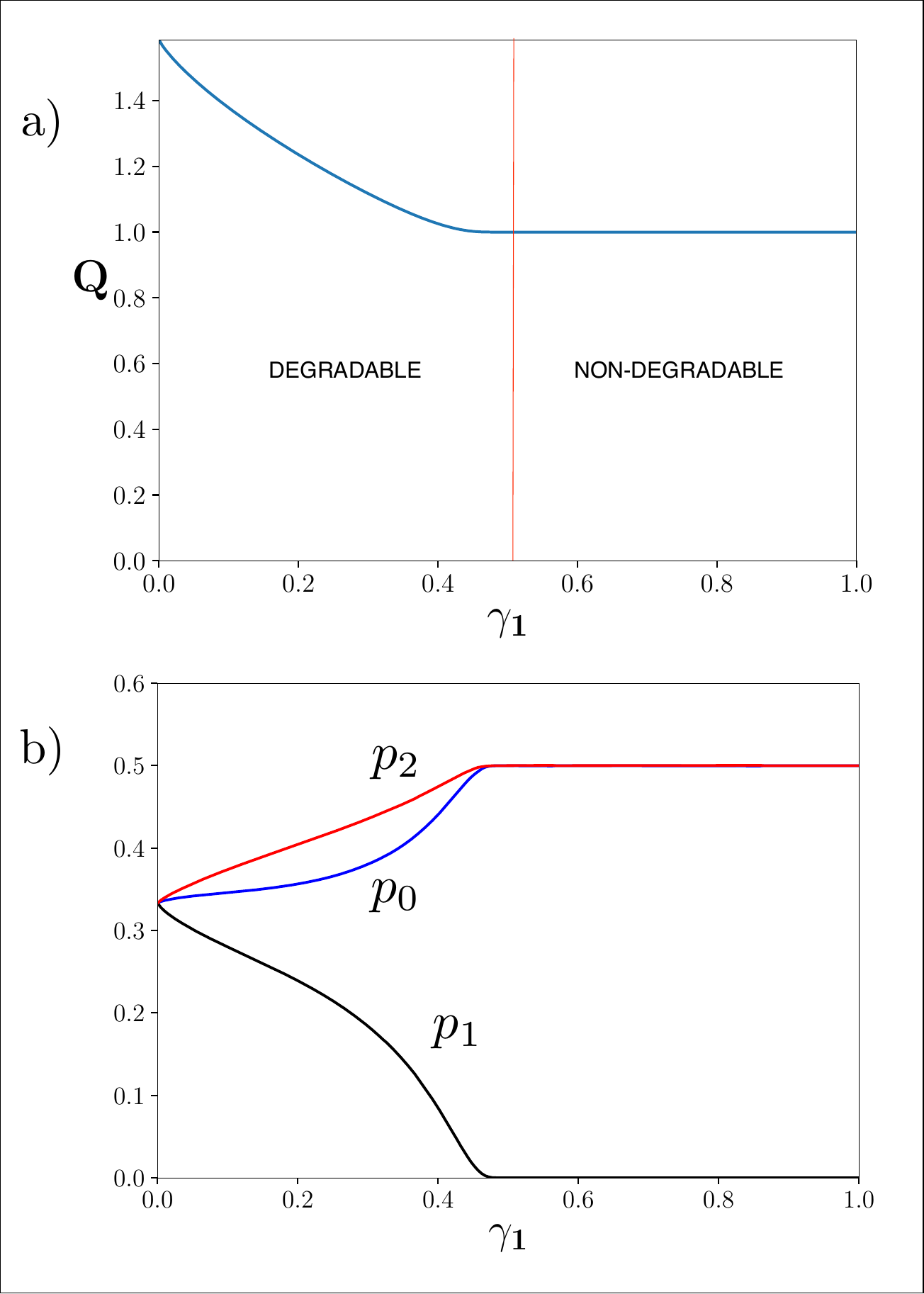}
  \caption{Upper panel: profile of the quantum  and the private classical capacity for the channel $\mathcal{D}_{(\gamma_1,0,0)}$ w.r.t. the damping parameter $\gamma_1$.
  For $\gamma_1\leq 1/2$ the channel is degradable and the reported value follows from the numerical solution of Eq.~(\ref{solveTHIS}). 
  For $\gamma>1/2$ instead the channel is neither degradable nor antidegradable: here the associated capacity value is equal to $1$ (see main text).
  Notice that the reported values respect the monotonicity property~(\ref{DECGM}). Lower panel: 
  populations $p_0$, $p_1$ and $p_2$ of those states that maximize the quantum capacity formula~(\ref{solveTHIS}) for the channel $\mathcal{D}_{(\gamma_1,0,0)}$ w.r.t. the damping parameter $\gamma_1$.
  }  \label{fig:Qg1}
\end{figure}

Despite the fact that the channel is degradable only for $0\leq \gamma_1 \leq \frac{1}{2}$ and that we know that it's not anti-degradable, 
we can still compute the value of the capacity of $\mathcal{D}_{(\gamma_1,0,0)}$  showing that 
\begin{eqnarray}\label{NICEID} 
Q(\mathcal{D}_{(\gamma_1,0,0)})=C_p(\mathcal{D}_{(\gamma_1,0,0)}) = 1\;, \qquad \forall \gamma_1\geq 1/2. 
\end{eqnarray} 
This indeed is a direct consequence of the lower bound~(\ref{LOW1}), the fact that $Q(\mathcal{D}_{(\gamma_1,0,0)})$ and $C_p(\mathcal{D}_{(\gamma_1,0,0)})$  are non-increasing functions of $\gamma_1$ as explicitly shown in Eq.~(\ref{DECGM}), and of the fact that from Eq.~(\ref{solveTHIS}) we get $Q(\mathcal{D}_{(\gamma_1=1/2,0,0)})= C_p(\mathcal{D}_{(\gamma_1=1/2,0,0)})=1$ by direct evaluation.
Putting all this together we obtain  
\begin{equation}
1= Q(\mathcal{D}_{(\gamma_1=1/2,0,0)})\geq Q(\mathcal{D}_{(\gamma_1,0,0)})  \geq 1\;, \qquad \forall  \gamma_1\geq 1/2,
\end{equation} 
that implies~(\ref{NICEID}), the same conclusion of course holding true for $C_p(\mathcal{D}_{(\gamma_1,0,0)})$.
The results discussed above are summarized in the plot in Fig.~\ref{fig:Qg1}.

\subsection{Complete damping of the first excited state ($\gamma_1=1$)} \label{sec:first} 
\begin{figure}[t!]
\includegraphics[width=\linewidth]{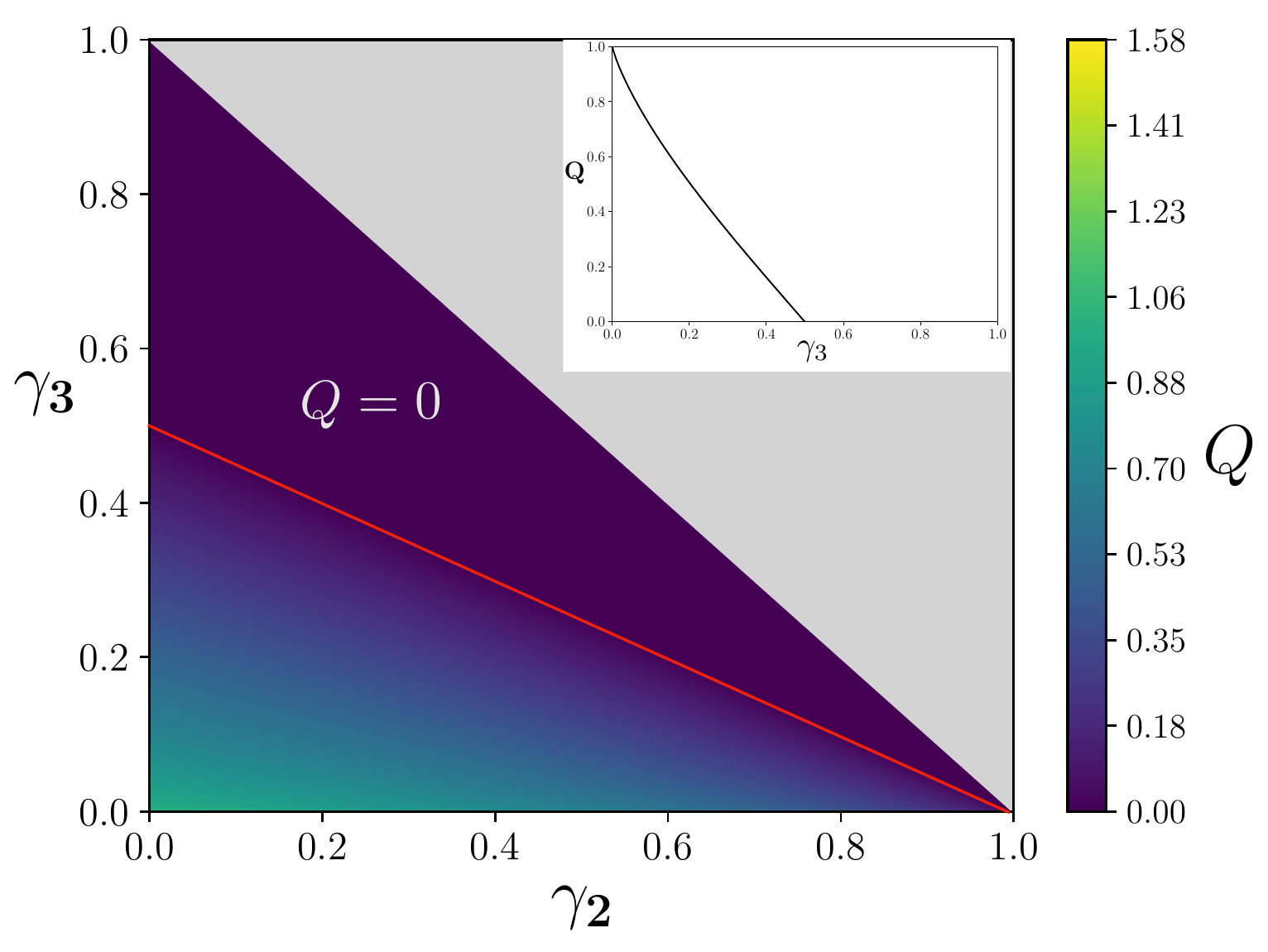}
  \caption{Quantum and private classical capacity of the MAD channel $\mathcal{D}_{(1,\gamma_2,\gamma_3)}$ w.r.t. $\gamma_2$ and $\gamma_3$, computed according to  Eq.~(\ref{exact}) -- the associated parameter region corresponds to the 
  ABC triangle of Fig.~\ref{fig:decay}. The grey region represent points where  $\mathcal{D}_{(1,\gamma_2,\gamma_3)}$ is not CPTP; the points above the red line ($\gamma_3= (1-\gamma_2)/2$)
  have zero capacity, $Q(\mathcal{D}_{(1,\gamma_2,\gamma_3)})=0$. The equivalent channel $\mathcal{D}'_{(\gamma_2,\gamma_3)}$ of Eq.~(\ref{eq:matrix form Phi1=123}) is anti-degradable for points above the red line and degradable below. For $\gamma_2=0$ the value of $Q(\mathcal{D}_{(1,\gamma_2,\gamma_3)})$ and $C_p(\mathcal{D}_{(1,\gamma_2,\gamma_3)})$
coincides with the quantum capacity~\cite{QUBIT ADC} of a qubit ADC channel of transmissivity $\gamma_3$ (see inset): this should be compared with
the value of $Q(\mathcal{D}_{(1,\gamma_2,\gamma_3)})$ and 
$C_p(\mathcal{D}_{(1,\gamma_2,\gamma_3)})$ on the other border (i.e. $\gamma_3=0$), which we report in Fig.~\ref{fig:Q1=12}. Notice finally that the reported values respect the monotonicity requirement of Eq.~(\ref{DECGMstrong}).}
  \label{fig:Q1=123}
\end{figure}

Assume next that our qutrit MAD channel of Eq.~(\ref{eq:matrix expression})  is characterized by the maximum value of $\gamma_1$ allowed by CPTP constraint of Eq.~(\ref{REGION}), i.e. 
 $\gamma_1=1$, region represented by the ABC triangle of Fig.~\ref{fig:decay}.
This map corresponds to the case where the  initial population of the first excited level $|1\rangle$,  gets completely lost in favor of the ground state $|0\rangle$ of the model 
so that  Eqs.~(\ref{eq:matrix expression}), (\ref{eq:matr expr compl}) rewrite as
\begin{widetext} 
\begin{eqnarray}\label{eq:matr expr 1=123}
\mathcal{D}_{(1,\gamma_2,\gamma_3)}(\hat{\rho})&=&\begin{pmatrix}
1-(1-\gamma_3) \rho_{22}& 0 & \sqrt{1-\gamma_2-\gamma_3}\rho_{02}\\
0 & \gamma_2 \rho_{22} & 0\\
\sqrt{1-\gamma_2-\gamma_3}\rho_{02}^* & 0 & (1-\gamma_2-\gamma_3)\rho_{22}
\end{pmatrix},\\ \nonumber \\ 
\tilde{\mathcal{D}}_{(1,\gamma_2,\gamma_3)}(\hat{\rho})
&=&\left(
\begin{array}{cccc}
\rho_{00}+ (1-\gamma_2-\gamma_3 ) \rho_{22} &  \rho_{01} & 0 & \sqrt{\gamma_3} \rho_{02} \\
 \rho_{01}^* & \rho_{11} & 0 &  \sqrt{\gamma_3} \rho_{12} \\
 0 & 0 & \gamma_2 \rho_{22} & 0 \\
 \sqrt{\gamma_3} \rho_{02}^* &  \sqrt{\gamma_3} \rho_{12}^* & 0 & \gamma_3 \rho_{22} \\
\end{array}
\right), \label{eq:matr expr 1=123ewe}
\end{eqnarray}
\end{widetext}
for $\gamma_2,\gamma_3\in[0,1]$ such that $\gamma_2+ \gamma_3\leq 1$. 
The above expressions make it explicit that, at variance with the case  discussed in
the previous section and in agreement with the conclusions of Ref.~\cite{PROP DEGR},
 the map $\mathcal{D}_{(1,\gamma_2,\gamma_3)}$ is not degradable. 
Indeed we notice that while  $\tilde{\mathcal{D}}_{(1,\gamma_2,\gamma_3)}(\hat{\rho})$ preserves information about the
components $\rho_{11}$, $\rho_{01}$, $\rho_{10}$, $\rho_{12}$, $\rho_{21}$ of the input state $\hat{\rho}$, no trace of those terms is left  in 
$\mathcal{D}_{(1,\gamma_2,\gamma_3)}(\hat{\rho})$: accordingly it is technically impossible to identify a linear  (not mentioning
CPTP) map ${\cal N}$ which applied to $\mathcal{D}_{(1,\gamma_2,\gamma_3)}(\hat{\rho})$ would reproduce 
$\tilde{\mathcal{D}}_{(1,\gamma_2,\gamma_3)}(\hat{\rho})$ for all $\hat{\rho}$. 
 Despite this fact it turns out  that
 also for $\mathcal{D}_{(1,\gamma_2,\gamma_3)}$, the capacity can still be expressed as the single letter expression~(\ref{eq:QCapacity1GasDIAG3}).
 Specifically, as we shall see in the following, in this case we can write  
 \begin{eqnarray}
Q(\mathcal{D}_{(1,\gamma_2,\gamma_3)}) &=&C_p(\mathcal{D}_{(1,\gamma_2,\gamma_3)})\nonumber \\
& =& Q^{(1)}(\mathcal{D}_{(1,\gamma_2,\gamma_3)}) ={\cal Q}(\gamma_2,\gamma_3) \;, \label{exact}
\end{eqnarray} 
with the function ${\cal Q}(\gamma_2,\gamma_3)$
being formally  defined as 
\begin{widetext} 
\begin{eqnarray}\nonumber 
{\cal Q}(\gamma_2,\gamma_3)&\equiv&
\max_{\hat{\tau}_{\text{diag}}} \left\{ S(\mathcal{D}_{(1,\gamma_2,\gamma_3)}(\hat{\tau}_{\text{diag}})) -
 S(\tilde{\mathcal{D}}_{(1,\gamma_2,\gamma_3)}(\hat{\tau}_{\text{diag}}))\right\}  
 \\
 &=&\max_{p\in [0,1]} \Big\{  - (1-(1-\gamma_3) p) \log_2(1-(1-\gamma_3) p) - (1-\gamma_2-\gamma_3)p \log_2 (1-\gamma_2-\gamma_3)p)\nonumber \\
&& + ( 1-(\gamma_2+\gamma_3) p) \log_2( 1-(\gamma_2+\gamma_3) p) +\gamma_3 p \log_2 \gamma_3 p)\Big\}\;, \label{eq:QCapacity12312}
\end{eqnarray}
\end{widetext} 
where the maximization is restricted to the diagonal density matrices $\hat{\tau}_{\text{diag}}=(1-p) |0\rangle\langle 0|+ p |2\rangle\langle 2|$  of $\text{A}'$, associated with the linear subspace ${\cal H}_{\text{A}'}\equiv \mbox{Span}\{\ket{0},\ket{2}\}$.
The explicit value of ${\cal Q}(\gamma_2,\gamma_3)$ has been numerically plotted in Fig.~\ref{fig:Q1=123}: 
we remark here that for  $\gamma_3\geq \frac{1-\gamma_2}{2}$ this function assumes zero value, i.e. ${\cal Q}(\gamma_2,\gamma_3) = 0$,
in agreement with the fact  that in such regime the channel $\mathcal{D}_{(1,\gamma_2,\gamma_3)}$ has zero capacity, i.e. 
 \begin{eqnarray}
Q(\mathcal{D}_{(1,\gamma_2,\gamma_3)}) &=&C_p(\mathcal{D}_{(1,\gamma_2,\gamma_3)})  = 0\;,  
\nonumber \\
&&   \forall \; 1-\gamma_2\geq  \gamma_3\geq \tfrac{1-\gamma_2}{2}\;.  \label{exact0}
\end{eqnarray}

To prove Eq.~(\ref{exact}) let us start by observing that ${\cal Q}(\gamma_2,\gamma_3)$ provides a natural lower bound for
$Q(\mathcal{D}_{(1,\gamma_2,\gamma_3)})$ and hence for $C_p(\mathcal{D}_{(1,\gamma_2,\gamma_3)})$: this is a simple consequence of~(\ref{eq:QCapacitylowe}),  which allows us to write 
\begin{eqnarray} \label{lower1}
Q(\mathcal{D}_{(1,\gamma_2,\gamma_3)}) &\geq &  \max_{\hat{\rho}}  J(\mathcal{D}_{(1,\gamma_2,\gamma_3)}, \hat{\rho}) \nonumber \\
&\geq& \max_{\hat{\tau}_{\text{diag}}}  J(\mathcal{D}_{(1,\gamma_2,\gamma_3)}, \hat{\tau}_{\text{diag}}) = {\cal Q}(\gamma_2,\gamma_3)\;,
\nonumber 
\end{eqnarray} 
with $J$ being the coherent information functional~(\ref{eq:coherent info}). 
Next step is now to show that the function ${\cal Q}(\gamma_2,\gamma_3)$ provides also an upper bound for $Q(\mathcal{D}_{(1,\gamma_2,\gamma_3)})$:
we do this by constructing a new channel $\mathcal{D}'_{(\gamma_2,\gamma_3)}$ 
whose capacity is provably better than the capacity of  $\mathcal{D}_{(1,\gamma_2,\gamma_3)}$, i.e.
\begin{eqnarray} 
Q(\mathcal{D}_{(1,\gamma_2,\gamma_3)}) &\leq& Q(\mathcal{D}'_{(\gamma_2,\gamma_3)}) \;, \label{impo1}  \\
C_p(\mathcal{D}_{(1,\gamma_2,\gamma_3)}) &\leq& C_p(\mathcal{D}'_{(\gamma_2,\gamma_3)}) \;, \label{impo1cp} 
\end{eqnarray} 
and for which we can explicitly show that 
\begin{eqnarray} Q(\mathcal{D}'_{(\gamma_2,\gamma_3)})=C_p(\mathcal{D}'_{(\gamma_2,\gamma_3)}) = {\cal Q}(\gamma_2,\gamma_3)\;.\label{impo2}  \end{eqnarray}  
For this purpose notice
 that since the population of level $|1\rangle$ is washed away,  
the output produced by $\mathcal{D}_{(1,\gamma_2,\gamma_3)}$ can be simulated by  the CPTP map
$\mathcal{D}'_{(\gamma_2,\gamma_3)}: {\cal L}({\cal H}_{\text{A}'}) \rightarrow {\cal L}({\cal H}_{\text{A}}) $ operating on the two levels quantum system associated with the Hilbert space ${\cal H}_{\text{A}'}\equiv \mbox{Span}\{\ket{0},\ket{2}\}$,
and producing qutrit states of A as outputs. In particular defining $\hat{\tau}$ a generic density matrix on 
 ${\cal H}_{\text{A}'}$ we have
\begin{equation}\label{eq:matrix form Phi1=123}
\mathcal{D}'_{(\gamma_2,\gamma_3)}(\hat{\tau})=\begin{pmatrix}
1-(1-\gamma_3) \tau_{22}& 0 & \sqrt{1-\gamma_2-\gamma_3}\tau_{02}\\
0 & \gamma_2 \tau_{22} & 0\\
\sqrt{1-\gamma_2-\gamma_3}\tau_{02}^* & 0 & (1-\gamma_2-\gamma_3)\tau_{22}
\end{pmatrix}
\end{equation} 
with the corresponding complementary channel~(\ref{eq:app compl chan}) given by 
\begin{equation}  \label{eq:matrix form Phi1=123tilde}
\tilde{\mathcal{D}}'_{(\gamma_2,\gamma_3)}(\hat{\tau})=\left(
\begin{array}{ccc}
 1-(\gamma_2+\gamma_3)\tau_{22} & 0 & \sqrt{\gamma_3}\tau_{02}\\
 0 & \gamma_2 \tau_{22} & 0\\
 \sqrt{\gamma_3}\tau_{02}^* & 0 & \gamma_3\tau_{22}
\end{array}
\right),
\end{equation}
where for $i,j=0,2$ we set $\tau_{ij}\equiv \langle i| \hat{\tau} | j\rangle$.

The  reason why $\mathcal{D}'_{(\gamma_2,\gamma_3)}$
 fulfills the inequality (\ref{impo1})  is a direct consequence of the  
 fact that $\mathcal{D}_{(1,\gamma_2,\gamma_3)}$, while yielding the same outcomes of $\mathcal{D}'_{(\gamma_2,\gamma_3)}$, 
  is  also ``wasting" resources in the useless level $|1\rangle$. To formalize this, notice that we can write
 \begin{eqnarray} 
\mathcal{D}_{(1,\gamma_2,\gamma_3)} = \mathcal{D}'_{(\gamma_2,\gamma_3)} \circ {\cal A} \;, 
\end{eqnarray} 
 where $\mathcal{A}: {\cal L}({\cal H}_{\text{A}}) \rightarrow
{\cal L}({\cal H}_{\text{A}'})$ is  the CPTP transformation  which maps the input state of the qutrit  A to the qubit system $\text{A}'$
by completely erasing the level  $\ket{1}$ 
and moving its population to $\ket{0}$, i.e. 
\begin{eqnarray}\label{eq:matr expr 1=123adsf}
\mathcal{A}(\hat{\rho})&=&\begin{pmatrix}
\rho_{00} + \rho_{11} & \rho_{02} \\
\rho_{20} &\rho_{22} \\
\end{pmatrix},
\end{eqnarray}
 where $\rho_{ij} =\langle i | \hat{\rho} |j\rangle$ with $\hat{\rho} \in \mathfrak{S}({\cal H}_{\text{A}})$.
\begin{figure}[t!]
\includegraphics[width=\linewidth]{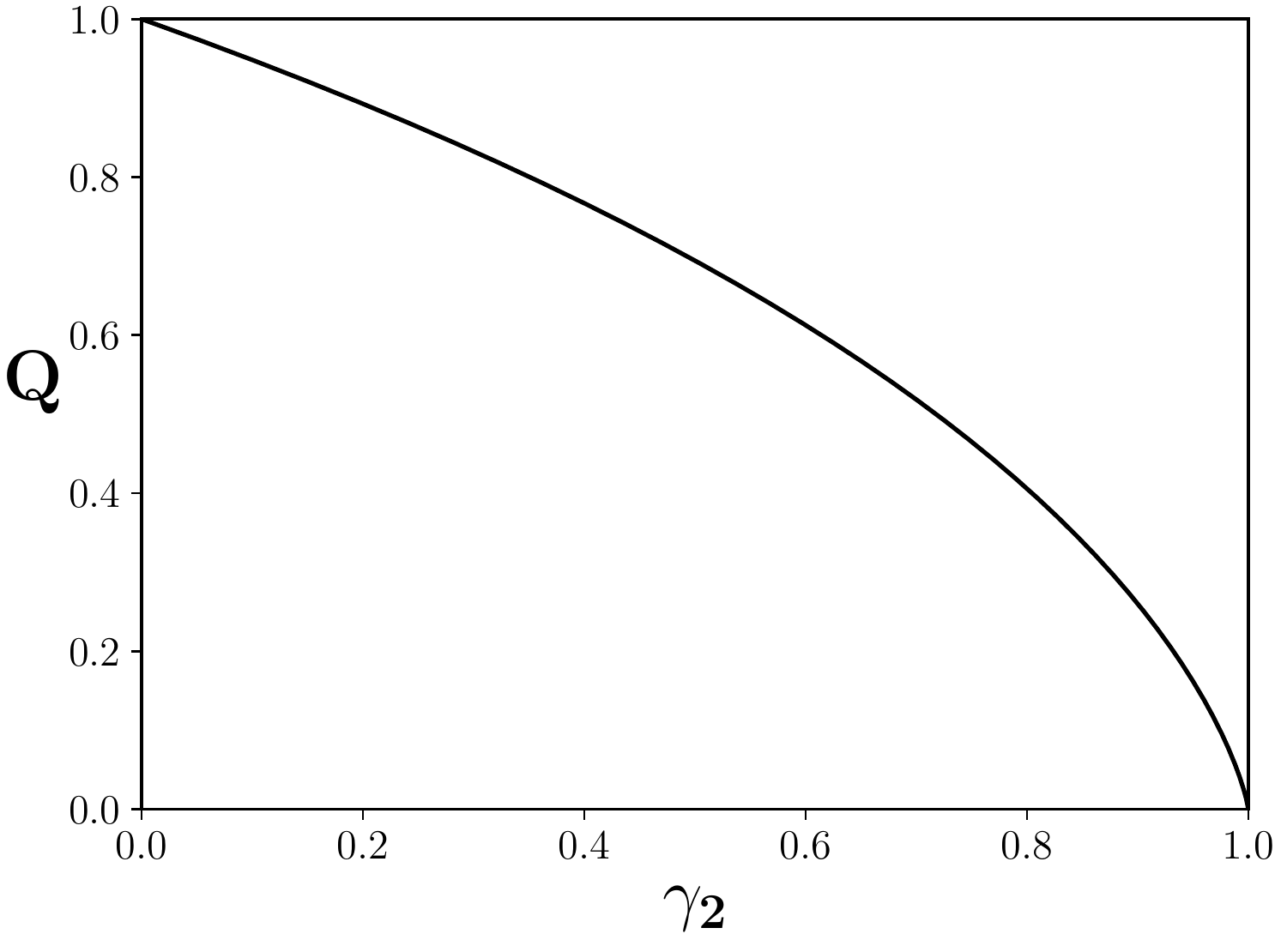}
  \caption{Quantum (and private classical) capacity of the channel $\mathcal{D}_{(1,\gamma_2,0)}$  w.r.t. $\gamma_2$.
Region corresponding to the edge AC of Fig.~\ref{fig:decay}. }
  \label{fig:Q1=12}
\end{figure}
  Equation~(\ref{impo1}) can hence be derived as a direct consequence of  the bottleneck inequality~(\ref{eq:data processing}) applied to the case in which 
  $\Gamma$ is indeed the quantum capacity $Q$. 
 The second part of the argument, i.e. Eq.~(\ref{impo2}), can instead be derived by noticing that  at variance with  the original mapping $\mathcal{D}_{(1,\gamma_2,\gamma_3)}$ which is never degradable, it turns out that 
$\mathcal{D}'_{(\gamma_2,\gamma_3)}$ is  degradable for 
\begin{eqnarray}0\leq  \gamma_3\leq ({1-\gamma_2})/{2}\;, \label{REDGED} \end{eqnarray}
and antidegradable otherwise, i.e. for  $(1-\gamma_2)/2\leq \gamma_3\leq {1-\gamma_2}$.
This can be shown for instance by observing  that in the region identified by the inequality~(\ref{REDGED}) 
the quantity 
\begin{eqnarray} \bar{\gamma}_3\equiv
 \frac{1-\gamma_2-2\gamma_3}{1-\gamma_2-\gamma_3}\;, \end{eqnarray} 
 belongs to the interval $[0,1]$ and can be used to build up a proper  CPTP single-decay qutrit MAD channel 
   $\mathcal{D}_{(0,0,\bar{\gamma}_3)}$ -- see Sec.~\ref{sec:single decay}. Furthermore 
  by direct calculation we also get  
  \begin{eqnarray}\label{deg} 
 \mathcal{D}_{(0,0,\bar{\gamma}_3)} \circ \mathcal{D}'_{(\gamma_2,\gamma_3)} = \tilde{\mathcal{D}}'_{(\gamma_2,\gamma_3)}\;,
 \end{eqnarray}  
 which shows that $\mathcal{D}_{(0,0,\bar{\gamma}_3)}$ acts as the 
 connecting channel
${\cal N}$ entering the degradability condition~(\ref{eq:degradable}) of $\mathcal{D}'_{(\gamma_2,\gamma_3)}$. 
   From Eqs.~(\ref{eq:matrix form Phi1=123}) and (\ref{eq:matrix form Phi1=123tilde}) 
 it is also immediately visible that $\mathcal{D}'_{(\gamma_2,\gamma_3)}$ can be obtained from $\tilde{\mathcal{D}}'_{(\gamma_2,\gamma_3)}$ by the substitution $\gamma_3\rightarrow 1-\gamma_2-\gamma_3$. Consequently using the same construction (\ref{deg}) we can conclude that 
 $\mathcal{D}'_{(\gamma_2,\gamma_3)}$  is antidegradable for $(1-\gamma_2)/2\leq \gamma_3\leq {1-\gamma_2}$.

To  derive Eq.~(\ref{impo2}) we finally observe that 
as the original mapping $\mathcal{D}_{(1,\gamma_2,\gamma_3)}$, also $\mathcal{D}'_{(\gamma_2,\gamma_3)}$ 
is covariant under the group of unitary transformations which are diagonal in the computational basis of the model: accordingly, following the same argument that
led us to (\ref{eq:QCapacity1GasDIAG3}), 
we can express its capacity as 
\begin{eqnarray} 
Q(\mathcal{D}'_{(\gamma_2,\gamma_3)})&=&C_p(\mathcal{D}'_{(\gamma_2,\gamma_3)})=\max_{\hat{\tau}_{\text{diag}}} \Big\{ S(\mathcal{D}'_{(\gamma_2,\gamma_3)}(\hat{\tau}_{\text{diag}})) \nonumber \\ \label{CAPQprime} 
&-&
 S(\tilde{\mathcal{D}}'_{(\gamma_2,\gamma_3)}(\hat{\tau}_{\text{diag}}))\Big\} = {\cal Q}(\gamma_2,\gamma_3),
\end{eqnarray}
 the last identity following from the fact that
  $\mathcal{D}'_{(\gamma_2,\gamma_3)}(\hat{\tau}_{\text{diag}}))$ coincides with 
   $\mathcal{D}_{(1,\gamma_2,\gamma_3)}(\hat{\tau}_{\text{diag}})$  and by the fact that 
 the  positive component of the spectrum of 
 $\tilde{\mathcal{D}}'_{(\gamma_2,\gamma_3)}(\hat{\tau}_{\text{diag}})$ coincides with the one of 
 $\tilde{\mathcal{D}}_{(1,\gamma_2,\gamma_3)}(\hat{\tau}_{\text{diag}})$ (strictly speaking the above derivation  holds true only in the
 degradable region~(\ref{REDGED}) of 
 $\mathcal{D}'_{(\gamma_2,\gamma_3)}(\hat{\tau}_{\text{diag}}))$: still since ${\cal Q}(\gamma_2,\gamma_3)$ nullifies for 
 $1-\gamma_2\geq \gamma_3\geq ({1-\gamma_2})/{2}$, we can apply~(\ref{CAPQprime}) also in the antidegradability region of the channel where
 $Q(\mathcal{D}'_{(\gamma_2,\gamma_3)})=0$). 
 
 As a concluding remark we comment on a special limit of the above construction obtained by setting  $\gamma_2=0$: in this case we notice that the effective map~(\ref{eq:matrix form Phi1=123})
 can be replaced with the quantum channel 
 \begin{equation}\label{eq:matrix form Phi1=123gamma2=0}
\mathcal{D}'_{\gamma_3}(\hat{\tau})=\begin{pmatrix}
1-(1-\gamma_3) \tau_{22}&  \sqrt{1-\gamma_3}\tau_{02}\\
\sqrt{1-\gamma_3}\tau_{02}^* & (1-\gamma_3)\tau_{22}
\end{pmatrix}\;, 
\end{equation}
which now maps the two-level system $\text{A}'$ into itself via a standard  qubit ADC map  with rate $\gamma_3$.
 Accordingly, following the same analysis we did before we can conclude that $Q(\mathcal{D}_{(1,0,\gamma_3)})$ coincides with the 
 capacity value of the latter, computed in Ref.~\cite{QUBIT ADC}.

\subsection{Double-decay qutrit MAD channel with $\gamma_2=0$}\label{sec:double decay}

\begin{figure}[]
  \includegraphics[width=\linewidth]{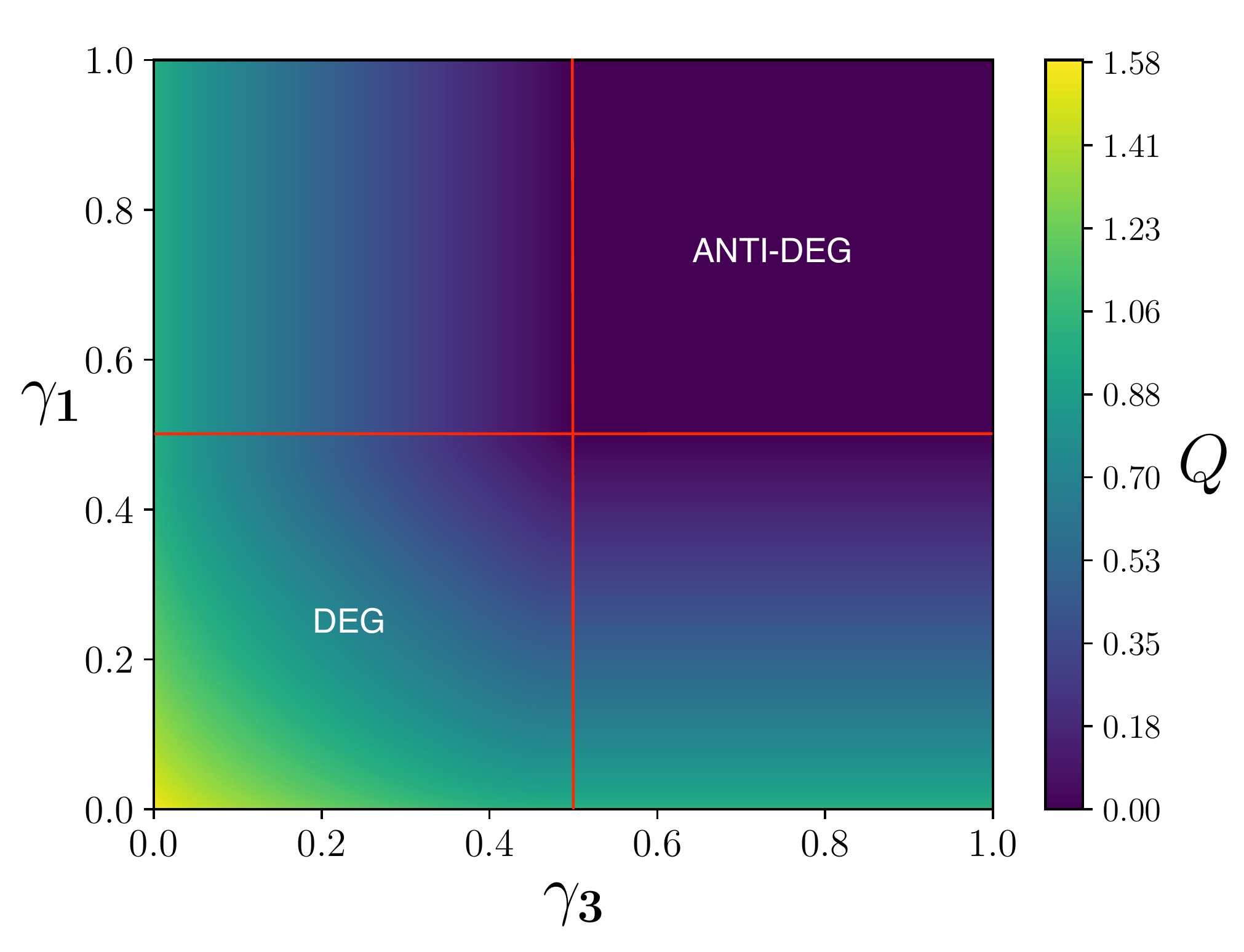}
  \caption{Quantum (and private classical) capacity for the channel $\mathcal{D}_{(\gamma_1,0,\gamma_3)}$ w.r.t. the damping parameters $\gamma_1$ and $\gamma_3$ -- square surface ABED of
Fig.~\ref{fig:decay}.
  For $\gamma_1,\gamma_2 \leq 1/2$ (region DEG), the channel is degradable and its capacity $Q$ is computed by solving numerically the maximization~(\ref{eq:cohID13});
  for $\gamma_1,\gamma_2 \geq 1/2$ (region ANTI-DEG) instead it is explicitly antidegradable and its capacity is zero. 
  Values in the SE and NW quadrants of the picture follow from the monotonicity behaviors~Eq.~(\ref{DECGMstrong}) and by the
  symmetry ~(\ref{SIMM2}): in particular in the SE
  sector the capacity is constant w.r.t. $\gamma_1$ (see Eq.~(\ref{GOODone})), while the NW is constant w.r.t. to $\gamma_3$. }
  \label{fig:Qg13_tot}
\end{figure}

\begin{figure}[t!]
\includegraphics[width=\linewidth]{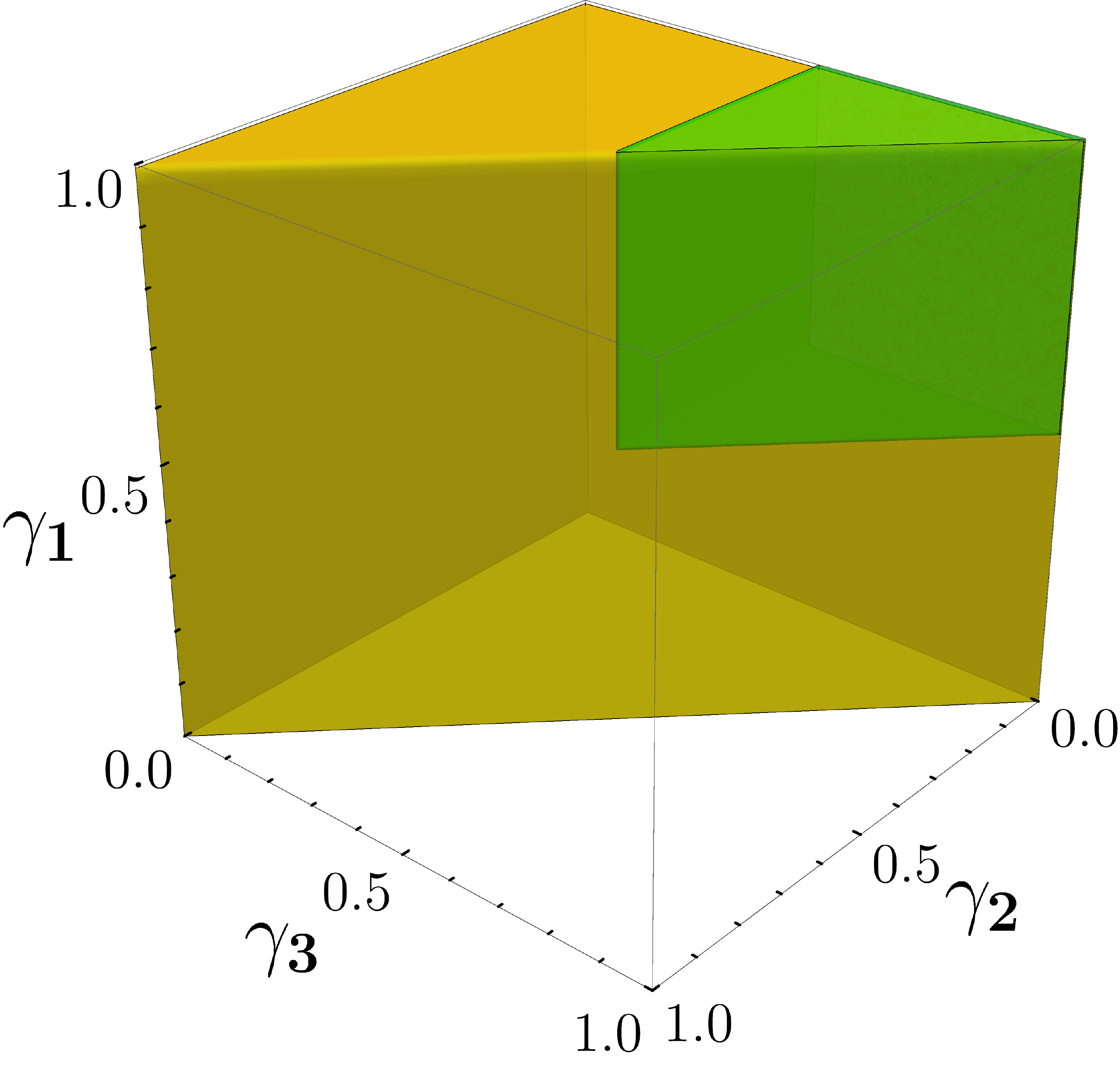}
  \caption{According to Eq.~(\ref{zerocap}) all points included in the green region of the plot have zero quantum  (and private classical) capacity. }
  \label{newfiggreen}
\end{figure}

Here we consider the value of the capacity for $\vec{\gamma}$ belonging to the square surface ABED of
Fig.~\ref{fig:decay}, identified by the condition $\gamma_2=0$.
From Eq.~(\ref{eq:Krausnewee}) we have that the Kraus operators for the MAD channel $\mathcal{D}_{(\gamma_1,0,\gamma_3)}$ are three:
\begin{eqnarray}\label{eq:Kraus13}
&&\hat{K}_0=\begin{pmatrix}
1 & 0 & 0\\
0 & \sqrt{1-\gamma_1} & 0 \nonumber \\
0 & 0 & \sqrt{1-\gamma_3}
\end{pmatrix}, \quad
\hat{K}_{01}=\begin{pmatrix}
0 & \sqrt{\gamma_1} & 0\\
0 & 0 & 0\\
0 & 0 & 0
\end{pmatrix}, \\
&&\qquad\qquad\quad\hat{K}_{03}=\begin{pmatrix}
0 & 0 & \sqrt{\gamma_3}\\
0 & 0 & 0\\
0 & 0 & 0
\end{pmatrix}
\end{eqnarray}
while  Eqs.~(\ref{eq:matrix expression}) and (\ref{eq:matr expr compl}) become 
\begin{widetext}
\begin{eqnarray}\label{eq:matrix form D13}
&\mathcal{D}_{(\gamma_1,0,\gamma_3)}(\hat{\rho})=\begin{pmatrix}
\rho_{00}+\gamma_1 \rho_{11}+\gamma_3 \rho_{22}& \sqrt{1-\gamma_1} \rho_{01} & \sqrt{1-\gamma_3}\rho_{02}\\
\sqrt{1-\gamma_1}\rho_{01}^* & (1-\gamma_1)\rho_{11} & \sqrt{1-\gamma_1}\sqrt{1-\gamma_3}\rho_{12}\\
\sqrt{1-\gamma_3}\rho_{02}^* & \sqrt{1-\gamma_1}\sqrt{1-\gamma_3}\rho_{12}^* & (1-\gamma_3)\rho_{22}
\end{pmatrix},\\ \label{eq:matrix form D13 tilde}
& \tilde{\mathcal{D}}_{(\gamma_1,0,\gamma_3)}(\hat{\rho})=\left(
\begin{array}{ccc}
 1-\gamma_1 \rho_{11}-\gamma_3 \rho_{22} & \sqrt{\gamma_1} \rho_{01} & \sqrt{\gamma_3} \rho_{02} \\
 \sqrt{\gamma_1} \rho_{01}^* & \gamma_1 \rho_{11} & \sqrt{\gamma_1} \sqrt{\gamma_3} \rho_{12} \\
 \sqrt{\gamma_3} \rho_{02}^* & \sqrt{\gamma_1} \sqrt{\gamma_3} \rho_{12}^* & \gamma_3 \rho_{22} \\
\end{array}
\right).
\end{eqnarray}
\end{widetext}
As evident from Fig.~\ref{BIGMA} and from the formal structure of Eq.~(\ref{eq:matrix form D13}),
for $\gamma_2=0$ the model exhibits a symmetry under the exchange of $\gamma_1$ and $\gamma_3$.
Indeed, indicating with $\hat{V}$ the unitary gate that swaps levels $|2\rangle$ and $|3\rangle$ we have that 
\begin{eqnarray}\label{EXC} 
\mathcal{D}_{(\gamma_3,0,\gamma_1)} (\hat{\rho}) = \hat{V} \mathcal{D}_{(\gamma_1,0,\gamma_3)} ( \hat{V} \hat{\rho} \hat{V}^\dag) \hat{V}^\dag\;,
\end{eqnarray} 
which by data-processing inequality implies 
\begin{eqnarray}\label{SIMM2} 
Q(\mathcal{D}_{(\gamma_1,0,\gamma_3)}) =Q(\mathcal{D}_{(\gamma_3,0,\gamma_1)}) \;,
\end{eqnarray} 
with an analogous identity applying in the case of the private classical capacity. 
Following  the procedure in Appendix \ref{sec:Appendix compl chan} we now observe that $\mathcal{D}_{(\gamma_1,0,\gamma_3)}$ is invertible 
for $\gamma_1,\gamma_3<1$, while $\tilde{\mathcal{D}}_{(\gamma_1,0,\gamma_3)}\circ\mathcal{D}_{(\gamma_1,0,\gamma_3)}^{-1}$ is CPTP for $\gamma_1,\gamma_3\leq \frac{1}{2}$, implying that in this range of parameters the channel is degradable (region DEG of  Fig.~\ref{fig:Qg13_tot}). 
Comparing  Eqs.~(\ref{eq:matrix form D13}) with (\ref{eq:matrix form D13 tilde}) we also realize that 
 \begin{eqnarray}
 \tilde{\mathcal{D}}_{({\gamma_1,0,\gamma_3})}=\mathcal{D}_{(1-\gamma_1,0,1-\gamma_3)}\;. \end{eqnarray}
 Therefore, by the same argument above, we can conclude that the channel is antidegradable for $\gamma_1,\gamma_3\geq \frac{1}{2}$ 
 (region ANTI-DEG of  Fig.~\ref{fig:Qg13_tot}) so that $Q(\mathcal{D}_{(\gamma_1,0,\gamma_3)})$ is null for that range of values.
Notice that resulting from Eq.~(\ref{eq:boundCOMBO1}) this translates to the following stronger statement:
\begin{equation} \label{zerocap} 
Q({\mathcal{D}}_{({\gamma_1,\gamma_2,\gamma_3})})=C_p({\mathcal{D}}_{({\gamma_1,\gamma_2,\gamma_3})}) =0 \;, \quad \forall \gamma_1,\gamma_3\geq \frac{1}{2}\;,
\end{equation}  
(see green region of Fig.~\ref{newfiggreen}).

To evaluate $Q({\mathcal{D}}_{({\gamma_1,0,\gamma_3})})$ and $C_p({\mathcal{D}}_{({\gamma_1,\gamma_2,\gamma_3})})$ in the region DEG of Fig.~\ref{fig:Qg13_tot}, where the map ${\mathcal{D}}_{({\gamma_1,0,\gamma_3})}$ is
provably degradable, we exploit Eq.~(\ref{eq:QCapacity1GasDIAG3}) obtaining 
\begin{widetext}
\begin{align}\label{eq:cohID13}
Q(&\mathcal{D}_{(\gamma_1,0,\gamma_3)})=C_p(\mathcal{D}_{(\gamma_1,0,\gamma_3)})
=\max_{p_1,p_2} \Big\{-[1 - (1-\gamma_1)p_1+(1-\gamma_3)p_2]\log_2[1 - (1-\gamma_1)p_1+(1-\gamma_3)p_2]\nonumber\\ 
&\qquad\qquad -(1-\gamma_1)p_1\log_2((1-\gamma_1)p_1) -(1-\gamma_3)p_2\log_2((1-\gamma_3)p_2)\nonumber\\ 
&\qquad\qquad +(1-\gamma_1p_1-\gamma_3p_2)\log_2(1-\gamma_1p_1-\gamma_3p_2) +\gamma_1 p_1\log_2(\gamma_1 p_1)+\gamma_3 p_2\log_2(\gamma_3 p_2)
\Big\},
\end{align}
\end{widetext} 
the maximization running over all possible values $p_1,p_2\in [0,1]$ under the constraint that $p_1+p_2\leq 1$.

 Notice that the capacities are known also on the borders of the parameters space, since when one of the rates is 0 we reduce to the single-decay MAD
 we solved in Sec.~\ref{sec:single decay}. 
  When one of the rate is instead 1 we reduce to the MAD channel discussed in Sec.~\ref{sec:first}, for which $Q$ is already available. More precisely in Sec.~\ref{sec:first} we computed $Q(\mathcal{D}_{(1,0,\gamma_3)})$, verifying that it coincides with the capacity of the qubit ADC: the value of 
   $Q(\mathcal{D}_{(\gamma_1,0,1)})$ follows from the latter via the symmetry~(\ref{SIMM2}). 
Since the value of $Q$  is available also on the borders of the DEG region, we can now compare $Q({\cal{D}}_{(\gamma_1,0,\gamma_3)})$ at $\gamma_3=\frac{1}{2}$ and $\gamma_3=1$, for all $\gamma_2\leq 1/2$. We find that the two are the same, i.e. 
$Q({\cal{D}}_{(\gamma_1,0,1)}) = Q({\cal{D}}_{(\gamma_1,0,1/2})$, 
Accordingly, invoking the monotonicity constraint (\ref{DECGMstrong}), we can finally conclude that
\begin{eqnarray}  \label{GOODone} 
Q({\cal{D}}_{(\gamma_1,0,1)}) = Q({\cal{D}}_{(\gamma_1,0,\gamma_3})\; \qquad \forall \gamma_3\;,
\end{eqnarray} 
which invoking the symmetry (\ref{SIMM2}) allows us to 
 evaluate the quantum capacity on the entire parameters region, see Fig.~\ref{fig:Qg13_tot}.

\subsection{Double-decay qutrit MAD channel with $\gamma_1=0$}\label{sec:double decay1}
Here we consider the triangular surface DEF of Fig.~\ref{fig:decay}.
From Eq.~(\ref{eq.Kraus}) we have that the Kraus operators for the MAD channel $\mathcal{D}_{(0,\gamma_2,\gamma_3)}$ are three:
\begin{eqnarray}\label{eq:Kraus23}
& &\hat{K}_0=\begin{pmatrix}
1 & 0 & 0\\
0 & 0 & 0\\
0 & 0 & \sqrt{1-\gamma_2-\gamma_3}
\end{pmatrix}\quad
\hat{K}_{12}=\begin{pmatrix}
0 & 0 & 0\\
0 & 0 & \sqrt{\gamma_2}\\
0 & 0 & 0
\end{pmatrix} \nonumber \\
& & \qquad\qquad\quad\hat{K}_{03}=\begin{pmatrix}
0 & 0 & \sqrt{\gamma_3}\\
0 & 0 & 0\\
0 & 0 & 0
\end{pmatrix}.
\end{eqnarray}
The actions of $\mathcal{D}_{(0,\gamma_2,\gamma_3)}$ and its complementary counterpart $\tilde{\mathcal{D}}_{(0,\gamma_2,\gamma_3)}$ on a generic density matrix $\hat{\rho}$ can hence be
described as
\begin{widetext}
\begin{eqnarray} \label{defgamma23} 
&\mathcal{D}_{(0,\gamma_2,\gamma_3)}(\hat{\rho})=\begin{pmatrix}
\rho_{00}+\gamma_3 \rho_{22} & \rho_{01} & \sqrt{1-\gamma_2-\gamma_3}\rho_{02}\\
\rho_{01}^* & \rho_{11}+\gamma_2 \rho_{22} & \sqrt{1-\gamma_2-\gamma_3}\rho_{12}\\
\sqrt{1-\gamma_2-\gamma_3}\rho_{02}^* & \sqrt{1-\gamma_2-\gamma_3}\rho_{12}^* & (1-\gamma_2-\gamma_3)\rho_{22}
\end{pmatrix},\\ \label{deftildegamma23} 
& \tilde{\mathcal{D}}_{(0,\gamma_2,\gamma_3)}(\hat{\rho})=\left(
\begin{array}{ccc}
 1-(\gamma_2+\gamma_3) \rho_{22} & \sqrt{\gamma_2} \rho_{12} & \sqrt{\gamma_3} \rho_{02} \\
  \sqrt{\gamma_2} \rho_{12}^* & \gamma_2 \rho_{22} & 0 \\
 \sqrt{\gamma_3} \rho_{02}^* & 0 & \gamma_3 \rho_{22} \\
\end{array}
\right),
\end{eqnarray}
\end{widetext}
 (notice that in this case, 
 differently of what happens with $\mathcal{D}_{(\gamma_1,0, \gamma_3)}$, the complementary channel is not an element of the MAD
 set). 
 By close inspection of Eq.~(\ref{defgamma23}), and as intuitively suggested by Fig.~\ref{fig:decay}, also these channels exhibit a symmetry 
 analogous to the one reported in Eq.~(\ref{EXC}), but this time with $\hat{V}$ being the swap operation exchanging levels $|0\rangle$ and $|1\rangle$, 
 which gives us 
 \begin{eqnarray}\label{SIMM2} 
Q(\mathcal{D}_{(0,\gamma_2,\gamma_3)}) =Q(\mathcal{D}_{(0,\gamma_3,\gamma_2)}) \;,
\end{eqnarray}
and an analogous identity for the private classical capacity.
Furthermore, as in the case of the single-decay qutrit MAD channel $\mathcal{D}_{(0,\gamma_2,0)}$, we notice that  $\mathcal{D}_{(0,\gamma_2,\gamma_3)}$ has a noiseless subspace, given here by $\{\ket{0},\ket{1}\}$, and we
can establish the following lower bound: 
\begin{eqnarray} 
 C_p(\mathcal{D}_{(0,\gamma_2,\gamma_3)}) \geq Q(\mathcal{D}_{(0,\gamma_2,\gamma_3)})\geq \log_2(2)=1\;. \label{CONST23} 
\end{eqnarray} 
In particular this tells us that $\mathcal{D}_{(0,\gamma_2,\gamma_3)}$ cannot be antidegradable (the same conclusion can be obtained by noticing that \cite{PROP DEGR} the map $\tilde{\mathcal{D}}_{(\gamma_2,0,\gamma_3)}$ has a
  kernel  that cannot be included into the kernel set of ${\mathcal{D}}_{(\gamma_2,0,\gamma_3)}$ -- e.g. the former
  contains $\ket{0}\!\!\bra{1}$ while the latter does not). 

\begin{figure}[t!]
\includegraphics[width=\linewidth]{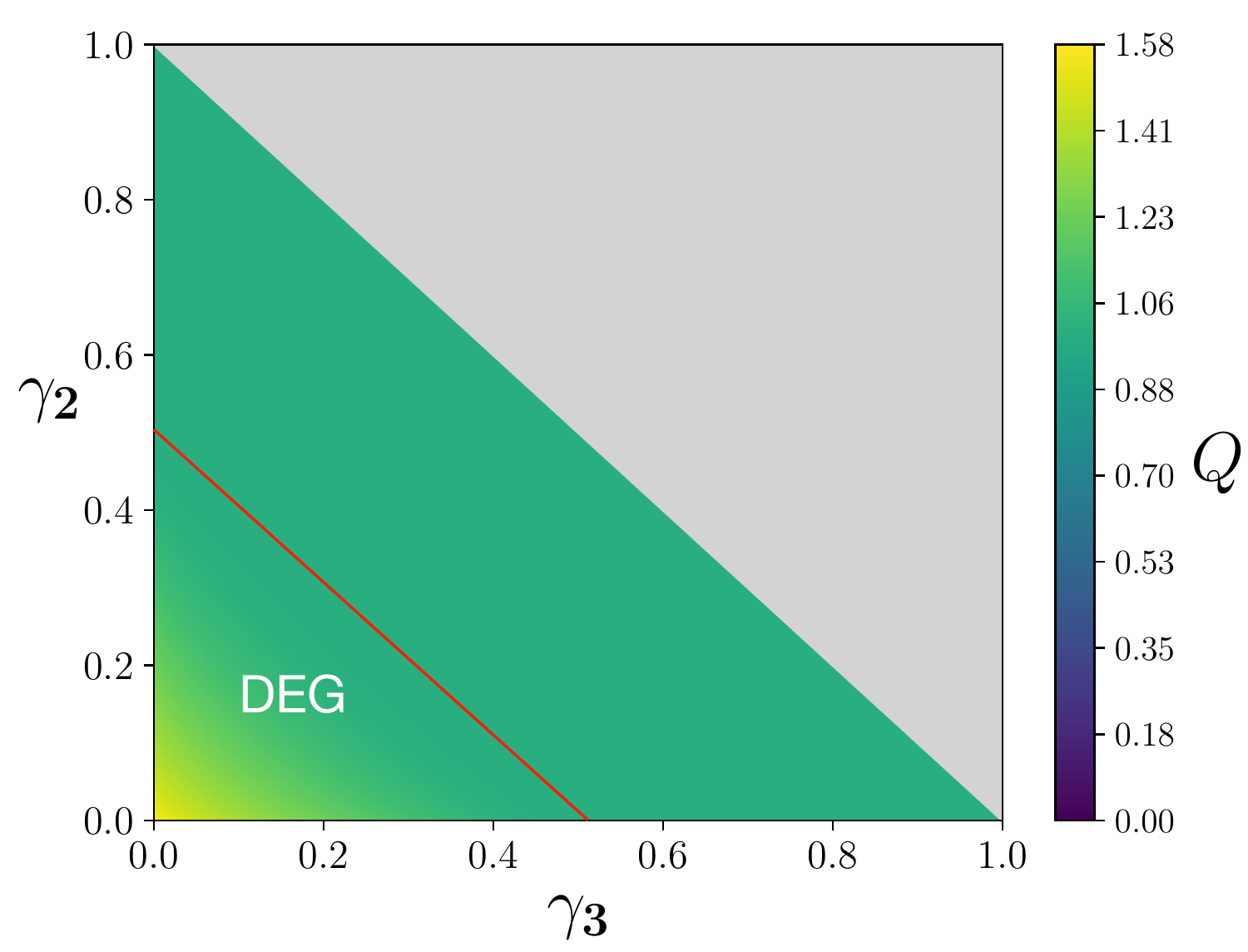}
  \caption{Quantum (and private classical) capacity of the channel $\mathcal{D}_{(0,\gamma_2,\gamma_3)}$ w.r.t. $\gamma_2$ 
  and $\gamma_3$ -- triangular surface DEF of Fig.~\ref{fig:decay}.
  The DEG zone below the red curve,  $\gamma_2+\gamma_3=\frac{1}{2}$, is the degradability region for the channel: here we compute $Q(\mathcal{D}_{(0,\gamma_2,\gamma_3)})$ 
  solving numerically the maximization of Eq.~(\ref{eq:IcohD23}). 
  Above the red curve the channel capacity assumes constant value~(\ref{CONST23}). 
  Notice that the reported function exhibits the symmetry (\ref{SIMM2}) and the monotonicity conditions~(\ref{DECGMstrong}).
  The grey zone indicates the non-accessible region~(\ref{REGION}). }
  \label{fig:23tot}
\end{figure}   

Following the usual approach we find that $\mathcal{D}_{(0,\gamma_2,\gamma_3)}$ 
is invertible for $\gamma_2+\gamma_3<1$, 
and that $\tilde{\mathcal{D}}_{(0,\gamma_2,\gamma_3)}\circ \mathcal{D}_{(0,\gamma_2,\gamma_3)}^{-1}$ is CPTP for $\gamma_2+\gamma_3\leq \frac{1}{2}$,
 which defines hence the degradability region for the map.   
 So, invoking (\ref{eq:QCapacity1GasDIAG3}) we  compute the quantum capacity in the degradability region as
\begin{equation}\label{eq:IcohD23}
\begin{split}
Q(&\mathcal{D}_{(0,\gamma_2,\gamma_3)})=C_p(\mathcal{D}_{(0,\gamma_2,\gamma_3)}) \\
&\quad\quad=\max_{p_0,p_1} \Big\{ -(p_1 +\gamma_2p_2)\log_2(p_1 +\gamma_2p_2)\\
&\quad\quad -[1-p_1-(1-\gamma_3) p_2]\log_2[1-p_1-(1-\gamma_3) p_2]\\
&\quad\quad -(1-\gamma_2-\gamma_3)p_2\log_2((1-\gamma_2-\gamma_3)p_2)\\
&\quad\quad +(1-(\gamma_2+\gamma_3)p_2)\log_2(1-(\gamma_2+\gamma_3)p_2)\\
&\quad\quad +\gamma_2 p_2\log_2(\gamma_2 p_2)+\gamma_3 p_2\log_2( \gamma_3 p_2)
\Big\}.
\end{split}
\end{equation}

Via numerical inspection we are also able to evaluate the magnitude of $Q$ on the border of the degradability region, 
designated by $\gamma_2+\gamma_3=\frac{1}{2}$, showing that here it equals the lower bound~(\ref{CONST23}). 
 This, in addition to the monotonicity~(\ref{DECGMstrong}), allows us to conclude that $\mathcal{D}_{(0,\gamma_2,\gamma_3)}$
 assumes the value 1 over all the region above the degradability borderline (red curve of Fig.~\ref{fig:23tot}), i.e. 
 \begin{eqnarray} 
Q(\mathcal{D}_{(0,\gamma_2,\gamma_3)})&=&C_p(\mathcal{D}_{(0,\gamma_2,\gamma_3)})=1\;, \nonumber 
\\
&& \qquad   \forall  \gamma_2+\gamma_3\geq 1/2\;. \label{CONST23up} 
\end{eqnarray}

\subsection{The qutrit MAD channel on the  $\gamma_2+\gamma_3=1$ plane}\label{sec:double decayplane}

Let us now consider the regime with $\gamma_2+\gamma_3=1$ where rate vectors $\vec{\gamma}$ belong to the rectangular area BEFC of Fig.~\ref{fig:decay}.

\begin{figure}[h!]
  \includegraphics[width=\linewidth]{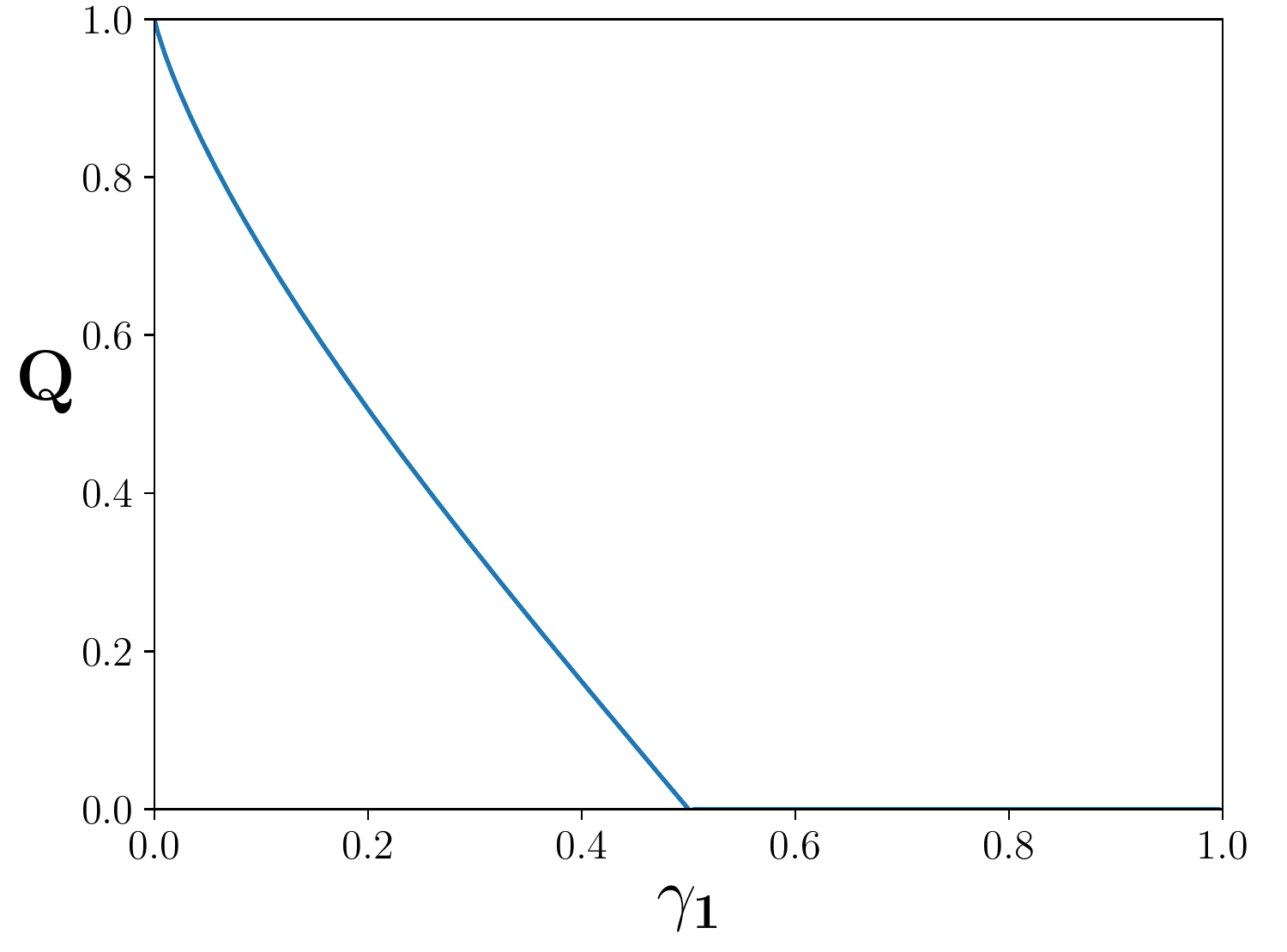}
  \caption{ Evaluation of $Q(\mathcal{D}_{(\gamma_1,\gamma_2,1-\gamma_2)})$ w.r.t. $\gamma_1$, equivalent to the qubit ADC quantum capacity (i.e. the rectangular region BEFC of Fig.~\ref{fig:decay}):
 as shown in Eq.~(\ref{Q1new}) the capacity exhibits no dependence upon $\gamma_2$ in this case.  }
  \label{fig:Q_qubit_ADC}
\end{figure}

Under this condition the map~(\ref{eq:matrix expression}) still admits four Kraus operators and  becomes  
\begin{equation}\label{eq:matrix expressionupper}
{\tiny{\mathcal{D}_{(\gamma_1,\gamma_2,1-\gamma_2)}(\hat{\rho})=\begin{pmatrix}
\rho_{00}+\gamma_1 \rho_{11}+(1-\gamma_2) \rho_{22}& \sqrt{1-\gamma_1}\rho_{01} &0\\
\sqrt{1-\gamma_1}\rho_{01}^* & (1-\gamma_1)\rho_{11}+\gamma_2 \rho_{22} & 0 \\
0& 0 & 0
\end{pmatrix}.}}
\end{equation}
We notice that the level $|2\rangle$ gets completely depopulated and that the channel can be expressed as 
\begin{eqnarray} 
\mathcal{D}_{(\gamma_1,\gamma_2,1-\gamma_2)} =  {\cal C}  \circ \mathcal{D}_{\gamma_1} \;,
\end{eqnarray} 
where  $\mathcal{D}_{\gamma_1}$ is a standard 
  qubit ADC channel connecting level $|1\rangle$ to level $|0\rangle$ with damping rate~$\gamma_1$, while now $\mathcal{C}$ is a CPTP transformation sending the qutrit A to the qubit system 
 spanned by vectors~$|0\rangle, |1\rangle$ and completely erasing the level  $\ket{2}$ , moving its population in part to $\ket{1}$ and in part to $\ket{0}$, i.e. 
\begin{eqnarray}\label{eq:matr expr 1=123adsfC}
\mathcal{C}(\hat{\rho})&=&\begin{pmatrix}
\rho_{00} +(1-\gamma_2) \rho_{22}  & \rho_{01} \\
\rho_{10} & \rho_{11} + \gamma_2 \rho_{22} \\
\end{pmatrix}.
\end{eqnarray}
Accordingly  the quantum capacity  of  $\mathcal{D}_{\gamma_1}$ computed in Ref.~\cite{QUBIT ADC} 
 is an explicit 
upper bound for $Q(\mathcal{D}_{(\gamma_1,\gamma_2,1-\gamma_2)})$ and $C_p(\mathcal{D}_{(\gamma_1,\gamma_2,1-\gamma_2)})$
(remember that for the qubit ADC $Q$ and $C_p$ coincide). 
On the other hand, $Q(\mathcal{D}_{\gamma_1})$ is also a lower bound for $Q(\mathcal{D}_{(\gamma_1,\gamma_2,1-\gamma_2)})$
and $C_p(\mathcal{D}_{(\gamma_1,\gamma_2,1-\gamma_2)})$
as its
rate can be achieved by simply using input states of A that live on the subspace $\{ |0\rangle, |1\rangle\}$. 
Consequently we can conclude that the following identity holds true
\begin{eqnarray}  \label{Q1new} 
Q(\mathcal{D}_{(\gamma_1,\gamma_2,1-\gamma_2)}) = C_p(\mathcal{D}_{(\gamma_1,\gamma_2,1-\gamma_2)}) =Q(\mathcal{D}_{\gamma_1}) \;,
\end{eqnarray} 
as shown in Fig.~\ref{fig:Q_qubit_ADC}.

\subsection{Double-decay qutrit MAD channel with $\gamma_3=0$}\label{sec:double decay12}
Here we consider the square region CADF of Fig.~\ref{fig:decay} identified by $\gamma_3=0$.
From Eq.~(\ref{eq.Kraus}) we have that the Kraus operators for $\mathcal{D}_{(\gamma_1,\gamma_2,0)}$ are three:
\begin{align}\label{eq:Kraus12}
\hat{K}_0=&\begin{pmatrix}
1 & 0 & 0\\
0 & \sqrt{1-\gamma_1} & 0\\
0 & 0 & \sqrt{1-\gamma_2}
\end{pmatrix}\quad
\hat{K}_{01}=\begin{pmatrix}
0 & \sqrt{\gamma_1} & 0\\
0 & 0 & 0\\
0 & 0 & 0
\end{pmatrix}\nonumber\\
&\qquad\qquad\quad\hat{K}_{02}=\begin{pmatrix}
0 & 0 & 0\\
0 & 0 & \sqrt{\gamma_2}\\
0 & 0 & 0
\end{pmatrix},
\end{align}
while the actions of $\mathcal{D}_{(\gamma_1,\gamma_2,0)}$ and $\tilde{\mathcal{D}}_{(\gamma_1,\gamma_2,0)}$ on a generic density matrix $\hat{\rho}$ are:
\begin{widetext}
\begin{eqnarray}\label{eq:matrix form D12}
&\mathcal{D}_{(\gamma_1,\gamma_2,0)}(\hat{\rho})=\begin{pmatrix}
\rho_{00}+\gamma_1 \rho_{11}& \sqrt{1-\gamma_1}\rho_{01} & \sqrt{1-\gamma_2}\rho_{02}\\
\sqrt{1-\gamma_1}\rho_{01}^* & (1-\gamma_1)\rho_{11}+\gamma_2 \rho_{22} & \sqrt{1-\gamma_1}\sqrt{1-\gamma_2}\rho_{12}\\
\sqrt{1-\gamma_2}\rho_{02}^* & \sqrt{1-\gamma_1}\sqrt{1-\gamma_2}\rho_{12}^* & (1-\gamma_2)\rho_{22}
\end{pmatrix}, \\
& \tilde{\mathcal{D}}_{(\gamma_1,\gamma_2,0)}(\hat{\rho})=\left(
\begin{array}{ccc}
 1-\gamma_1\rho_{11}-\gamma_2\rho_{22} & \sqrt{\gamma_1} \rho_{01} & \sqrt{1-\gamma_1}\sqrt{\gamma_2} \rho_{02} \\
 \sqrt{\gamma_1} \rho_{01}^* & \gamma_1 \rho_{11} & 0 \\
 \sqrt{1-\gamma_1}\sqrt{\gamma_2} \rho_{02}^* & 0 & \gamma_2 \rho_{22} \\
\end{array}
\right).\label{eq:matrix form D12tilde}
\end{eqnarray}
\end{widetext}
At variance with the previous sections,
we have that while  ${\mathcal{D}_{(\gamma_1,\gamma_2,0)}}$ is invertible for $\gamma_1,\gamma_2 < 1$,
 for no range of these values the application ${\tilde{\mathcal{D}}_{(\gamma_1,\gamma_2,0)}}\circ{\mathcal{D}^{-1}_{(\gamma_1,\gamma_2,0)}}$ 
  produces a CPTP map. We can hence conclude that the map is never degradable. 
   About antidegradability, here also we have that
$\ker\{\tilde{\mathcal{D}}_{(\gamma_1,\gamma_2,0)}\}\nsubseteq \ker\{\mathcal{D}_{(\gamma_1,\gamma_2,0)}\}$,
so $\mathcal{D}_{(\gamma_1,\gamma_2,0)}$ is also not antidegradable~\cite{PROP DEGR}.
As a matter of fact the only cases for which we can produce explicit values of $Q({\mathcal{D}_{(\gamma_1,\gamma_2,0)}})$
are the limiting cases where either $\gamma_1$ or $\gamma_2$ equals  $0$ (in these cases the map is a single-rate MAD channel 
discussed in Sec.~\ref{sec:single decay}), or $1$  where  instead the results of Sec.~\ref{sec:first} or Sec.~\ref{sec:double decayplane}
can be applied. 
For the remaining cases we resort in presenting a lower bound
for $Q(\mathcal{D}_{(\gamma_1,\gamma_2,0)})$ and $C_p(\mathcal{D}_{(\gamma_1,\gamma_2,0)})$.
\begin{figure}[t!]
  \includegraphics[width=\linewidth]{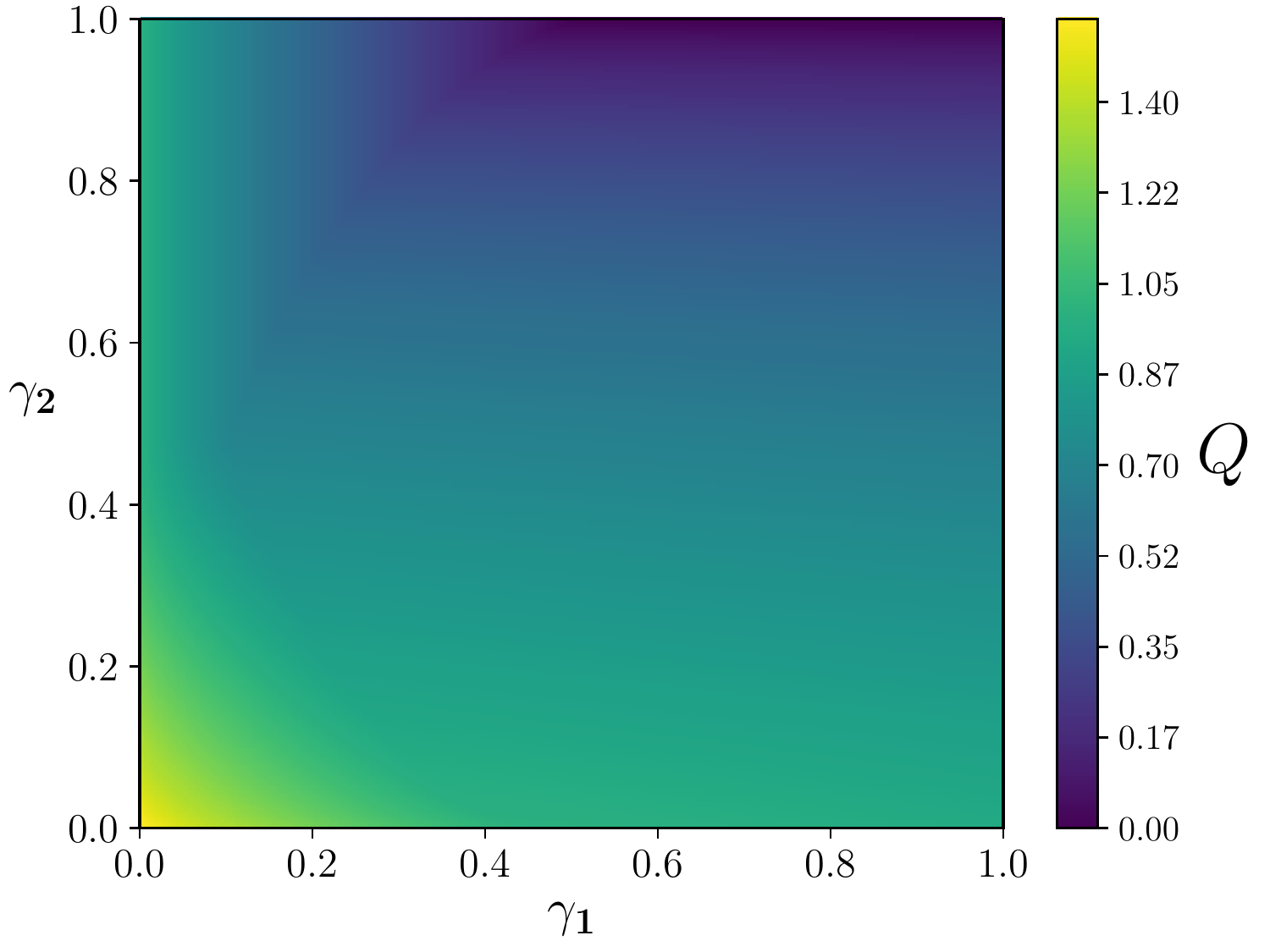}
  \caption{Numerical evaluation of a lower bound for $Q(\mathcal{D}_{(\gamma_1,\gamma_2,0)})$  (and $C_p(\mathcal{D}_{(\gamma_1,\gamma_2,0)})$) 
  obtained by maximizing the single use coherent information of the channel 
  over all possible diagonal inputs 
  -- the parameters region corresponds to the CADF square of Fig.~\ref{fig:decay}.
  Notice that reported plot does not fulfill the monotonicity constraint~(\ref{DECGMstrong}), hence
  explicitly proving that the function we present is certainly not the real capacity of the system.  }
  \label{fig:Q12_lb}
\end{figure}

A straightforward approach is to exploit the right-hand-side of Eq.~(\ref{eq:QCapacity1GasDIAG3})
and run them also outside the degradability region, in synthesis evaluating the maximum of the coherent information
of $\mathcal{D}_{(\gamma_1,\gamma_2,0)}$
on the diagonal sources. Notice that since the map is not degradable, the coherent information is not necessarily concave and the
restriction to diagonal sources does not even guarantee  that the computed expression corresponds to the true 
$Q^{(1)}(\mathcal{D}_{(\gamma_1,\gamma_2,0)})$ functional. 
Clearly the task can be refined as much as needed, e.g. by choosing less specific families of states or by computing $Q^{(i)}(\mathcal{D}_{(\gamma_1,\gamma_2,0)})$ for $i>1$, but these aspects are beyond the focus of this work and will be considered in future research. 
 The results we obtain are reported in Fig.~(\ref{fig:Q12_lb}).

\section{Entanglement Assisted Quantum Capacity of qutrit MAD channels}\label{sec:Ent ass}

\begin{figure}[t!]
  \includegraphics[width=\linewidth]{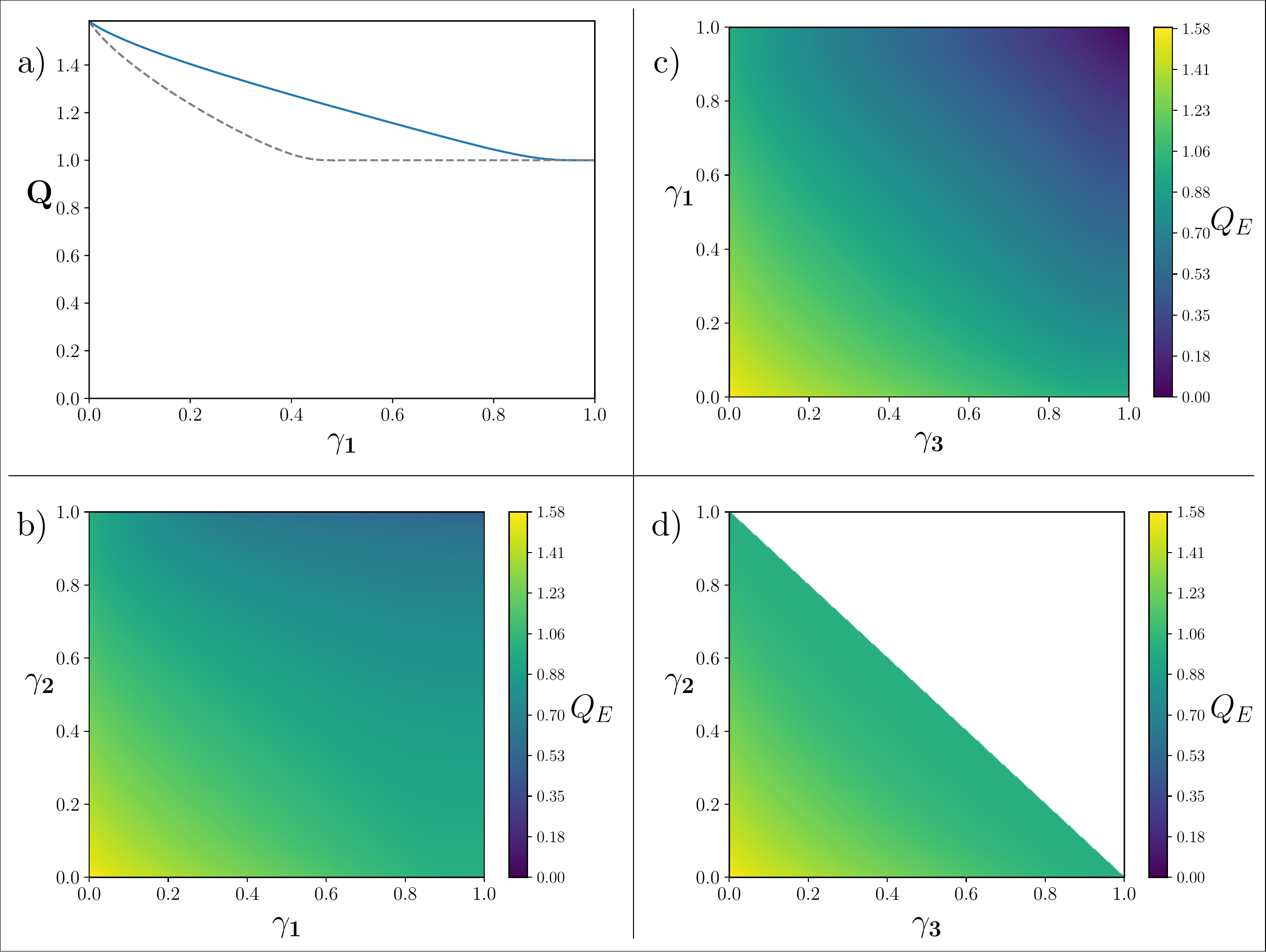}
  \caption{\textbf{a)} Profile of the entanglement assisted quantum capacity $Q_E(\mathcal{D}_{(\gamma_1,0,0)})$ w.r.t. the damping parameter $\gamma_1$ (results should be compared with those of Fig.~\ref{fig:Qg1} where we present
 $Q(\mathcal{D}_{(\gamma_1,0,0)})$ and $C_p(\mathcal{D}_{(\gamma_1,0,0)})$). Notice that also in this case the expression fulfills the monotonicity constraint~(\ref{DECGM}). In 
   \textbf{b)},  \textbf{c)}, \textbf{d)}  Entanglement assisted quantum capacity for the CADF square region of Fig.~\ref{fig:decay},  
    for the ABED region, and for the DEF region, respectively. 
  }
  \label{fig:Qe_merge}
\end{figure}

For the sake of completeness the present section is devoted to studying 
the entanglement assisted quantum capacity $Q_E(\mathcal{D})$ of MAD CPTP maps which quantifies the amount of quantum information transmittable per channel use assuming the communicating parties to share an arbitrary amount of entanglement. A general introduction to the subject is presented in Appendix~\ref{appQE} where 
 we review some basic properties and derive
 a simplified expression which in the case of 
  MAD channels of arbitrary dimension translates into 
 \begin{equation}
Q_{E}({\cal D})
=\frac{1}{2} \max_{\hat{\rho}_{\text{diag}}}\left\{  S(\hat{\rho}_{\text{diag}}) + S({\cal D}(\hat{\rho}_{\text{diag}})) -S(\tilde{\cal D}(\hat{\rho}_{\text{diag}})) \right\}\;, 
 \label{SIMPLIQE} 
\end{equation}
where $\hat{\rho}_{\text{diag}}$ are input density matrices which are diagonal in the computational basis of the system. 
 In the case of  the single-rate qutrit MAD transformations this translates to solving the following maximization: 
\begin{eqnarray}\label{eq:ent ass cap Dg1}
& &Q_E(\mathcal{D}_{(\gamma_1,0,0)}) =\frac{1}{2}\max_{p_0,p_1} \Big\{ -p_0\log_2p_0 - p_1\log_2p_1 \nonumber  \\
& &-2(p_0+\gamma_1 p_1)\log_2(p_0+\gamma_1 p_1) \nonumber \\
& &  -(1-\gamma_1)p_1\log_2((1-\gamma_1)p_1) \nonumber \\
&  &  +(1-\gamma_1p_1)\log_2(1-\gamma_1p_1)+\gamma_1p_1\log_2(\gamma_1 p_1)\Big\}, 
\end{eqnarray}
the result being reported in Fig.~\ref{fig:Qe_merge} a). In a similar fashion we also numerically compute $Q_E$ for all the two-rate qutrit MAD channels scenarios we analyzed in the previous sections, reporting the associated results in Fig.~\ref{fig:Qe_merge} b), c), d). Notice that also the three-rate qutrit MAD channels $Q_E$ can be computed but not easily visualized, hence it's not reported.

\section{Conclusions}\label{sec:conc}

We introduce a finite dimensional generalization of the qubit ADC model which represents one of the most studied examples of quantum noise
in quantum information theory. 
In this context the quantum (and private classical) capacity of a large class of quantum channels (namely the qutrit MAD channels) has been explicitly computed,
vastly extending the set of models whose capacity is known: : this effort in particular includes some non-trivial examples of quantum maps which are explicitly non-degradable (neither antidegradable) -- see e.g. the results of Sec.~\ref{sec:double decay}. 
Besides allowing  generalizations to higher dimensional systems (see e.g. Ref.~\cite{ARTICOLO1}), 
the analysis here presented naturally spawns further research, e.g. extending it to include other capacity measures, such as
the classical capacity or the two-way quantum capacity~\cite{WILDE1,PIR}.
We finally conclude by noticing that the MAD channel scheme discussed in the present paper can be also easily adapted to include generalizations of the (qubit) generalized amplitude damping channel scheme~\cite{WILDE1}, by 
 allowing reverse {\it damping} processes which promote excitations from lower to higher levels that could mimic, e.g., thermalization events.

We acknowledge support from PRIN 2017 ``Taming complexity with quantum strategies".

SC thanks P. Novelli for useful discussions.

\newpage

\twocolumngrid

\appendix

\section{Mathematical prerequisites} \label{MATHAPP}

Here we review some basic notions on quantum channels and quantum capacities that 
are extensively used in the main text. 
\subsection{Complementary channels and degradability}\label{sec:Appendix compl chan}
A CPTP map $\Phi:\mathcal{L}(\mathcal{H}_{\text{A}})\rightarrow \mathcal{L}(\mathcal{H}_{\text{B}})$ can be seen as the evolution induced by an isometry $\hat{V}:\mathcal{H}_{\text{A}}\rightarrow \mathcal{H}_{\text{B}}\otimes \mathcal{H}_{\text{E}}$ involving an environment $E$, called Stinespring dilation~\cite{CPTP,STINE}. Specifically for all input states $\rho_{\text{A}}
\in \mathfrak{S}_{\text{A}}$ we can write 
\begin{equation}
\Phi(\hat{\rho}_{\text{A}})=\text{Tr}_{\text{E}}[\hat{V}\hat{\rho}_A \hat{V}^{\dagger}].
\end{equation}
If instead we trace out the degrees of freedom in B we obtain the complementary (or conjugate) channel $\tilde{\Phi}:\mathcal{L}(\mathcal{H}_{\text{A}})\rightarrow \mathcal{L}(\mathcal{H}_{\text{E}})$, i.e. 
\begin{equation} \label{DEFCONJ} 
\tilde{\Phi}(\hat{\rho}_{\text{A}})=\text{Tr}_{\text{B}}[\hat{V}\hat{\rho}_{\text{A}} \hat{V}^{\dagger}]\;. 
\end{equation}
Being $\hat{M}_k$ the Kraus operators generating $\Phi$ and $\ket{k}_{\text{E}}$ a basis for the environment, the operator $\hat{V}$ can be written as:
\begin{equation}
\hat{V}=\sum_k \hat{M}_k \otimes \ket{k}_E,
\end{equation}
and being 
\begin{equation}
\hat{V}\hat{\rho}_{\text{A}} \hat{V}^{\dagger}=\sum_{i,j}  \hat{M}_i \hat{\rho}_{\text{A}} \hat{M_j}^{\dagger}\otimes \ket{i}_{\text{E}}\!\bra{j},
\end{equation}
it's straightforward to verify that Eq.~(\ref{DEFCONJ}) can be equivalently expressed as 
\begin{equation}\label{eq:app compl chan}
\tilde{\Phi}(\hat{\rho}_{\text{A}})=\sum_{i,j} \text{Tr}_{\text{B}}[ \hat{M}_i \hat{\rho}_{\text{A}} \hat{M_j}^{\dagger}]\ket{i}_{\text{E}}\!\bra{j}.
\end{equation}

 A fact that it is worth mentioning, as it  will play a fundamental role in our analysis, is that \cite{HOLEVO} for a channel $\Phi$ that is covariant under a unitary representation  of some  group $G$, i.e. 
 \begin{equation} \label{COVAR} 
 {\Phi}(\hat{U}^{\text{A}}_g \hat{\rho} \hat{U}_g^{\text{A}\dagger})=\hat{U}_g^{\text{B}}{\Phi}( \hat{\rho}) \hat{U}_g^{\text{B}\dagger}
 \;,\qquad \forall \hat{\rho}\in \mathfrak{S}({\cal H}), \forall g\in G\;, 
 \end{equation} then 
 also the complementary channel $\tilde{\Phi}$ is covariant under the same transformations, i.e.
\begin{equation}\label{eq:compl covariance} 
\tilde{\Phi}(\hat{U}^{\text{A}}_g \hat{\rho} \hat{U}_g^{\text{A}\dagger})=\hat{U}_g^{\text{E}}\tilde{\Phi}( \hat{\rho}) \hat{U}_g^{{\text{E}}\dagger}\;, \qquad \forall \hat{\rho}\in \mathfrak{S}({\cal H}), \forall g\in G\;,
\end{equation}
where  for X=A,B,E,  $\hat{U}_g^{\text{X}}$ is the unitary operator that represents the element $g$ of the group $G$ in the output space~X.

We finally recall the definition of degradable and anti-degradable channels~\cite{DEGRADABLE}. 
A quantum channel $\Phi$ is said degradable if a CPTP map $\mathcal{N}:\mathcal{L}(\mathcal{H}_{\text{B}})\rightarrow \mathcal{L}(\mathcal{H}_{\text{E}})$ exists s.t.
\begin{equation}\label{eq:degradable}
\tilde{\Phi}=\mathcal{N}\circ \Phi,
\end{equation}
while it's said antidegradable if it exists a CPTP map $\mathcal{M}:\mathcal{L}(\mathcal{H}_{\text{E}})\rightarrow \mathcal{L}(\mathcal{H}_{\text{B}})$ s.t.
\begin{equation}\label{eq:antidegradable}
\Phi=\mathcal{M}\circ\tilde{\Phi},
\end{equation}
(the symbol  ``$\circ$"  representing channel concatenation).
Notice that in case $\Phi$ is mathematically invertible, a simple direct way to determine whether it is degradable or not is to formally 
invert (\ref{eq:degradable}) constructing the super-operator $\tilde{\Phi}\circ \Phi^{-1}$ and check whether such object  is CPTP (e.g. by 
studying the positivity of its Choi matrix) \cite{INVERSE1,INVERSE}, i.e. 
explicitly 
\begin{equation} \label{NECANDSUFF} 
\mbox{$\Phi$ invertible} \Longrightarrow \mbox{$\Phi$ degradable iff $\tilde{\Phi}\circ \Phi^{-1}$ is CPTP} \;. 
\end{equation} 
 Concretely this can be done by using the fact that since quantum channels 
are linear maps connecting vector spaces of linear operators, 
they can in turn being represented as matrices acting on vector spaces. This through the following vectorization isomorphism:
\begin{eqnarray}
\hat{\rho}_{\text{A}}=\sum_{ij} \rho_{ij}\ket{i}_{\text{A}}\!\bra{j}&\longrightarrow&\ket{\rho\rangle}=\sum_{ij} \rho_{ij}\ket{i}_{\text{A}}\otimes \ket{j}_{\text{A}}\in 
{\cal H}_{\text{A}}^{\otimes 2} \nonumber \\
\label{eq:isomorphism}\\
\Phi(\hat{\rho}_{\text{A}})&\longrightarrow& \hat{M}_{\Phi} \ket{\rho\rangle}, \nonumber 
\end{eqnarray}
where now $\hat{M}_{\Phi}$ is a $d_{\text{B}}^2\times d_{\text{A}}^2$ matrix connecting ${\cal H}_{\text{A}}^{\otimes 2}$ and 
 ${\cal H}_{\text{B}}^{\otimes 2}$ ($d_{\text{A}}$ and $d_{\text{B}}$ being respectively the dimensions of ${\cal H}_{\text{A}}$ and ${\cal H}_{\text{B}}$),
 which given a Kraus set $\{\hat{M}_k\}_k$ for $\Phi$ it can be explicitly expressed as \begin{equation}
\hat{M}_{\Phi}=\sum_k \hat{M_k}\otimes \hat{M_k}^*.
\end{equation}
Following Eq.~(\ref{eq:degradable}) we have hence that  for a degradable channel the following identity must apply 
\begin{equation}
\hat{M}_{\tilde{\Phi}}=\hat{M}_{\mathcal{N}}\hat{M}_{\Phi},
\end{equation}
with $\hat{M}_{\mathcal{N}}$ the matrix representation of the CPTP connecting channel ${\cal N}$, 
implying that the super-operator $\tilde{\Phi}\circ \Phi^{-1}$ is now represented by matrix $\hat{M}_{\tilde{\Phi}}\hat{M}_{\Phi}^{-1}$.

\subsection{The quantum capacity of a quantum channel}\label{sec:Quant and Priv}

The quantum capacity $Q(\Phi)$  is a measure of how faithfully quantum states
can transit from the input to the output of the quantum channel $\Phi$ by exploiting proper  encoding and decoding
procedures that act on multiple transmission stages~\cite{HOLEVO BOOK,WILDE,WATROUSBOOK,HOLEGIOV}. 
A close, yet cumbersome, expression for $Q(\Phi)$ can be obtained 
in the form~\cite{QCAP1,QCAP2,QCAP3} 
\begin{equation}
\begin{split}\label{eq:QCapacity}
Q(\Phi)&
=\lim_{n\rightarrow \infty} \max
_{\hat{\rho}^{(n)}\in {\mathfrak{S}({\cal H}^{\otimes n})}}
\frac{1}{n} J(\Phi^{\otimes n},\hat{\rho}^{(n)}),
\end{split}
\end{equation}
where $\hat{\rho}^{(n)}$ is a generic joint density matrix belonging to the input Hilbert space $\mathcal{H}^{\otimes n}$ on which
the tensor extension $\Phi^{\otimes n}$ of $\Phi$ acts. 
The  quantity appearing in the right-hand-side of Eq.~(\ref{eq:QCapacity}) is the  coherent information functional
\begin{equation}\label{eq:coherent info}
J(\Phi^{\otimes n},\hat{\rho}^{(n)})\equiv S(\Phi^{\otimes n}(\hat{\rho}^{(n)}))-S(\tilde{\Phi}^{\otimes n}(\hat{\rho}^{(n)})),
\end{equation}
with  
 $S(\cdots) \equiv -\mbox{Tr}[ \cdots \log_2 \cdots]$ the von Neumann entropy~\cite{HOLEVO BOOK} and with $\tilde{\Phi}$ the complementary channel of $\Phi$ introduced in the previous
 section.

The expression in Eq.~(\ref{eq:QCapacity}) isn't in general easily computable due to the fact that the coherent
information functional is typically non sub-additive, making hard to take care of the 
 regularization limit on $n$: removing it will in general produce just a lower bound to $Q(\Phi)$, i.e. 
 \begin{equation}
\begin{split}\label{eq:QCapacitylowe}
Q(\Phi)&\geq Q^{(1)}(\Phi)\equiv \max_{\hat{\rho}\in {\mathfrak{S}({\cal H})}} J(\Phi,\hat{\rho})\;,
\end{split}
\end{equation}
where now the maximization is performed on all possible input states $\hat{\rho}$ of a single application of $\Phi$.
 Things however simplify a lot if $\Phi$ is antidegradable \cite{ANTIDEGRADABLE} or degradable \cite{DEGRADABLE}. Indeed in the first case one can invoke a no cloning argument to directly conclude that $Q(\Phi)=0$.
In the second case instead, the gap in Eq.~(\ref{eq:QCapacitylowe}) closes allowing us to compute $Q(\Phi)$ as 
\begin{equation}
\begin{split}\label{eq:QCapacity1}
Q(\Phi)& 
=Q^{(1)}(\Phi) \equiv \max_{\hat{\rho}\in {\mathfrak{S}({\cal H})}} J(\Phi,\hat{\rho}).
\end{split}
\end{equation}
Besides allowing for the single-letter simplification~(\ref{eq:QCapacity1}), 
another important consequence of the degradability property~(\ref{eq:degradable}) is the fact that, for channels fulfilling such condition,
the coherent information~(\ref{eq:coherent info}) is known to be concave~\cite{CONC,INVERSE1} with respect to the input 
state $\hat{\rho}$, i.e. 
\begin{eqnarray}
\label{eq:coherent infoconcave}
J(\Phi,\sum_k p_k \hat{\rho}_k) \geq \sum_k p_k J(\Phi, \hat{\rho}_k) \;, 
\end{eqnarray} 
for all statistical ensemble of input states $\{ p_k ; \hat{\rho}_k\}$.\
This last inequality  allows for some further drastic simplification in particular when the channel $\Phi$ is covariant under a group of unitary transformations
as in Eq.~(\ref{COVAR}). Indeed thanks to results in Ref.~\cite{HOLEVO} and the invariance of the von Neumann entropy under unitary operations we can now observe that
\begin{eqnarray}\label{eq:coherent infog}
J(\Phi,\hat{U}^{\text{A}}_g \hat{\rho} \hat{U}_g^{\text{A}\dagger}) &=&S({\Phi}(\hat{U}^{\text{A}}_g \hat{\rho} \hat{U}_g^{\text{A}\dagger})-S(\tilde{\Phi}(\hat{U}^{\text{A}}_g \hat{\rho} \hat{U}_g^{\text{A}\dagger})) \nonumber \\ \nonumber 
&=& S(\hat{U}_g^{\text{B}}{\Phi}( \hat{\rho}) \hat{U}_g^{\text{B}\dagger})-S(\hat{U}_g^{\text{E}}\tilde{\Phi}( \hat{\rho}) \hat{U}_g^{{\text{E}}\dagger}) 
\nonumber \\
&=& J(\Phi, \hat{\rho})\;, 
\end{eqnarray}
for all input states and for all elements $g$ of the group.
Given then a generic input state $\hat{\rho}$ of the system, construct the following ensemble
of density matrices  $\{ d\mu(g) ; \hat{\rho}_g\}$ with $d\mu(g)$ some properly defined  probability distribution on $G$ and 
with $\hat{\rho}_g \equiv \hat{U}_g^{\text{A}} \hat{\rho} \hat{U}_g^{\text{A}\dag}$. Defining then 
\begin{equation}\label{ave}
\Lambda_G[{\hat{\rho}}] \equiv \int d\mu_g \hat{\rho}_g = \int d\mu_g \hat{U}_g^{\text{A}}  \hat{\rho} \hat{U}_g^{\text{A}\dag} \;, 
\end{equation}
the average state of $\{ d\mu(g) ; \hat{\rho}_g\}$  we notice that if $\Phi$ is degradable 
the following inequality holds true: 
\begin{equation}
\label{eq:Icoh tilde}
J(\Phi,\Lambda_G[{\hat{\rho}}]) 
\geq \int d\mu_g  J(\Phi, \hat{U}_g^{\text{A}}  \hat{\rho} \hat{U}_g^{\text{A}\dag}) = J(\Phi, \hat{\rho} ) \;,
\end{equation}
where in the last passage we used the invariance~(\ref{eq:coherent infog}).
Accordingly we can now restrict the maximization in Eq.~(\ref{eq:QCapacity1}) to only those input states $\hat{\rho}_G$ which 
result from the averaging operation~(\ref{ave}), i.e. 
\begin{equation}
\begin{split}\label{eq:QCapacity1G}
Q(\Phi)= Q^{(1)}(\Phi)
=\max_{\hat{\rho}_G} J(\Phi,\hat{\rho}_G).
\end{split}
\end{equation}
For the special case of the MAD channels ${\cal D}$ introduced in Sec.~\ref{sec.Model}, thanks to Eq.~(\ref{COV}) we can identify the group $G$ 
with the set of unitary operations which are diagonal in the computational basis $\{|i\rangle\}_{i=0, \cdots, d-1}$.
Taking $d\mu_g$ a flat measure, Eq.~(\ref{ave}) allows us to identify $\Lambda_G[{\hat{\rho}}]$ with the
density matrices of A which are diagonal as well, i.e. 
\begin{equation}\label{aveDIAG}
\Lambda_G[{\hat{\rho}}]  = \mbox{diag}[{\hat{\rho}}]  \;,
\end{equation}
and therefore to derive from (\ref{eq:QCapacity1G}) the following compact expression: 
\begin{equation}
\begin{split}\label{eq:QCapacity1GDIAGO}
Q({\cal D})= Q^{(1)}({\cal D})
=\max_{\hat{\rho}_{\text{diag}}} J({\cal D},\hat{\rho}_{\text{diag}}),
\end{split}
\end{equation}
which for $d_{\text{C}}=3$ reduces to Eq.~(\ref{eq:QCapacity1GasDIAG3}) of the main text. 
For completeness we report also an alternative, possibly more explicit way to derive~(\ref{eq:QCapacity1GDIAGO}).
This is obtained by
observing that a special instance of the unitaries  which are diagonal in the computational basis of a MAD channel and hence fulfill the identity~(\ref{COV}), is provided by the  subgroup  $\mathcal{O}_D(d)$ formed by  the operators represented by the diagonal 
 $d\times d$ matrices for which  all the non-zero (and diagonal) elements are $\pm 1$.
  Clearly the identity operator $\hat{\mathds{1}}$ is an element of $\mathcal{O}_D(d)$
   and the group is finite with $2^d$ elements. Given then an arbitrary input state $\hat{\rho}$ of A, construct 
   then the ensemble $\{ p_k ; \hat{\rho}_k\}$ formed by the density matrices $\hat{\rho}_k\equiv \hat{O}_k \hat{\rho} 
   \hat{O}_k^\dag$, with $\hat{O}_k$ being the $k$-th element of $\mathcal{O}_D(d)$, and by a flat probability set 
   $p_k= 1/2^d$. 
It can be shown \cite{STACK} that the average state of $\{ p_k ; \hat{\rho}_k\}$ is diagonal in the computational basis, i.e.  
\begin{equation}\label{eq:diag ensemble}
 \frac{1}{2^d}\sum_{k=0}^{2^{d}-1} \hat{O}_k \hat{\rho} \hat{O}^\dagger_k=\text{diag}(\hat{\rho})\;,
\end{equation}
from which (\ref{eq:QCapacity1GDIAGO}) can once more be derived as a consequence of (\ref{eq:QCapacity1G})
for all degradable ${\cal D}$. 

\subsection{Private Classical Capacity}\label{appCP} 
The private classical capacity $C_p(\Phi)$ of a quantum channel  $\Phi$ quantifies the amount of information that the sender  and the receiver of the messages can exchange privately, i.e. without a third party able to extract information from the communication line. 
This quantity provides a natural upper bound for $Q(\Phi)$, i.e. 
\begin{eqnarray} 
Q(\Phi) \leq C_p(\Phi)\;,
\end{eqnarray} 
and a closed formula for it is given in \cite{QCAP3, PRIV2}:
\begin{equation}
C_p(\Phi)=\lim_{n\to \infty} \frac{1}{n} C_p^{(1)}(\Phi^{\otimes n}).
\end{equation}
where now, given a generic quantum ensemble $\mathcal{E}:=\{p_i, \hat{\rho}_i\}$ at the input of the channel $\Phi$, the one-shot expression
$C_p^{(1)}(\Phi)$ is computed as  
\begin{equation}
C_p^{(1)}(\Phi)=\max_{\mathcal{E}}\left(\chi(\Phi,\mathcal{E})-\chi(\tilde{\Phi},\mathcal{E})\right),
\end{equation} 
with 
\begin{eqnarray} 
\chi(\Phi,\mathcal{E})\equiv S(\Phi(\sum_i p_i \hat{\rho}_i)) - \sum_i p_i S(\Phi(\hat{\rho}_i)\;, 
\end{eqnarray} 
 the Holevo information~\cite{HOLEVO BOOK,WILDE,WATROUSBOOK} of the ensemble ${\cal E}$ computed at the output of the channel $\Phi$. 
Since $\chi(\Phi,\mathcal{E})$ 
 is not additive \cite{PRIV3}, the relation between the one-shot formula and the asymptotic formula is not trivial, making the computation of the latter difficult in general. Nonetheless  if the channel considered is degradable or antidegradable the task of finding the regularized private classical capacity simplifies~\cite{PRIV4}: indeed for degradable maps $\Phi$  we have 
\begin{equation}
C_p(\Phi)=Q(\Phi)= Q^{(1)}(\Phi).
\end{equation}
while for anti-degradable maps one has
$C_p(\Phi)=Q(\Phi)=Q^{(1)}(\Phi) =0$.

\subsection{Entanglement assisted quantum capacity} \label{appQE} 
The entanglement assisted quantum capacity $Q_E(\Phi)$ of the quantum channel $\Phi$ quantifies the amount of quantum information transmittable per channel use assuming the communicating parties to share an arbitrary amount of entanglement. A closed expression for it has been provided in Ref.~ \cite{ENT ASS1, ENT ASS2} and results in an expression which, in contrast to the quantum capacity formula, doesn't need a regularization w.r.t. to the number of channel uses, i.e. 
\begin{equation}\label{eq:ent ass cap}
Q_{E}(\Phi)
=\frac{1}{2} \max_{\hat{\rho}\in {\mathfrak{S}({\cal H})}} I(\Phi,\hat{\rho}),
\end{equation}
where now 
\begin{eqnarray}
I(\Phi,\hat{\rho}) &\equiv& S(\hat{\rho}) +J(\Phi,\hat{\rho})  \nonumber \\
&=& S(\hat{\rho}) + S(\Phi(\hat{\rho})) - S(
\tilde{\Phi}(\hat{\rho}))\;, 
\end{eqnarray} 
is the quantum mutual information functional. As in the case of $C_p(\Phi)$, $Q_{E}(\Phi)$ provides a natural upper bound for $Q(\Phi)$. 
 
 We remind that $I(\Phi,\hat{\rho})$ is concave in the input state \cite{WILDE}, i.e. 
 \begin{eqnarray}
I(\Phi,\sum_k p_k \hat{\rho}_k) &\geq& \sum_k p_k I(\Phi, \hat{\rho}_k) \;,
\end{eqnarray}
 for all ensembles $\{ p_k, \hat{\rho}_k\}$. 
 Exploiting this fact, in case the channel $\Phi$ is covariant under the action of some group of unitary transformations as in Eq.~(\ref{COVAR}), we can hence follow the same derivation detailed at the end of the previous section to claim that 
 \begin{equation}
\begin{split}\label{eq:QCapacity1GE}
Q_E(\Phi)
=\frac{1}{2} \max_{\hat{\rho}_G} I(\Phi,\hat{\rho}_G),
\end{split}
\end{equation}
where now we can restrict the maximization in Eq.~(\ref{eq:ent ass cap}) to only those input states $\hat{\rho}_G$ which 
result from the averaging operation~(\ref{ave}). Applying this to the covariance~(\ref{COV}) of MAD channels with respect to the unitary
transformations which are diagonal in the computational basis finally yields to
Eq.~(\ref{SIMPLIQE}) of the main text.

\end{document}